**Title:** Accounting for spatial confounding in epidemiological studies with individual-level exposures: An exposure-penalized spline approach

**Short Title:** Accounting for spatial confounding via exposure-penalized splines

**Authors:** Jennifer F. Bobb[1,2], Maricela F. Cruz[1], Stephen J. Mooney[3], Adam Drewnowski[3], David Arterburn[1], Andrea J. Cook[1,2]

**Author Affiliations**:

1. Kaiser Permanente Washington Health Research Institute, Seattle, WA
2. University of Washington Department of Biostatistics, Seattle, WA
3. University of Washington Department of Epidemiology, Seattle, WA

**Corresponding Author:** Jennifer F. Bobb; 1730 Minor Avenue, Suite 1600, Seattle, WA, 98101; Jennifer.f.bobb@kp.org

**Summary:**

In the presence of unmeasured spatial confounding, spatial models may actually increase (rather than decrease) bias, leading to uncertainty as to how they should be applied in practice. We evaluated spatial modeling approaches through simulation and application to a big data electronic health record study. Whereas the risk of bias was high for purely spatial exposures (e.g., built environment), we found very limited potential for increased bias for individual-level exposures that cluster spatially (e.g., smoking status). We also proposed a novel exposure-penalized spline approach that selects the degree of spatial smoothing to explain spatial variability in the exposure. This approach appeared promising for efficiently reducing spatial confounding bias.

# 1. Introduction

Unmeasured confounding poses a critical threat to observational studies (Rothman, Greenland, and Lash 2008). In epidemiological studies with outcomes measured at an individual level, confounding can occur at multiple levels, including individual, temporal, and spatial. The level at which confounding occurs depends on the sources of variability in the exposure. For spatial-level exposures such as measures of ambient air pollution or the built environment, confounding arises only at the spatial level due to exposure variability across individuals in different locations (spatial confounding). For individual-level exposures such as smoking status or medication use, confounding can arise due to both spatially varying factors patterned with exposure (such as education or socioeconomic status [SES], which cluster spatially) and individual-level factors (such as biological sex) that vary among individuals living in the same location.

One approach to address spatial confounding is to fit a spatial model. This is commonly done by including a spatially correlated random effect or by adjusting for fixed effect-terms of spatial location. Although spatial modeling is conceptually straightforward, prior statistical literature has shown that it can actually increase bias relative to fitting a non-spatial model, even when there is an important unmeasured spatial confounder (Hodges and Reich 2010; Paciorek 2010; Page et al. 2017; Clayton, Bernardinelli, and Montomoli 1993; Schnell and Papadogeorgou 2020). Hodges and Reich (2010) illustrated this potential for increased bias, which can be attributed to collinearity of the spatial term with the exposure of interest, highlighting that merely including a spatial random effect in the regression model is not a substitute for adjusting for unmeasured spatial confounding.

Paciorek (2010) characterized a range of scenarios in which fitting a spatial model has the potential to increase rather than decrease bias. The scenarios differed in the spatial scales of the confounded and unconfounded variability in the exposure, with spatial scale referring to the distance at which variation occurs, such as fine-scale neighborhood-level or large-scale regional variability. Paciorek (2010) showed that fitting a spatial model can increase bias when the spatial scale of the confounded variability in the exposure is smaller than the spatial scale of the unconfounded variability. However, in the discussion the author comments that the situation where bias is increased "may be of limited practical interest, because it's not clear that there are real applications in which the unconfounded variation in the exposure occurs at larger scales than the confounded variation" (Paciorek 2010). Given the potential for increased bias along with uncertainty around whether a given application is in a scenario of increasing bias, it is unclear whether standard spatial modeling approaches should be applied routinely in practice in the presence of unmeasured spatial confounding.

These prior studies that investigated the impact of spatial modeling on reducing (or increasing) bias due to unmeasured spatial confounding focused on scenarios in which the exposure variable of interest varied solely spatially. In epidemiologic practice, there are two types of exposures that vary spatially: (1) purely spatial exposures that smoothly vary over space but do not vary across individuals within the same location (e.g., built environment measures, ambient air pollution); and (2) exposures or risk factors that are specific to an individual (e.g., smoking status, medication use, blood pressure) but also vary spatially. In this work, we build on prior work regarding spatial confounding by distinguishing between these types of exposure data. We propose a framework in which the exposure of interest can be decomposed into a spatial component and a non-spatial component, and we conduct simulation studies in which we vary the relative contributions of these components. In addition, motivated by this exposure decomposition framework, we propose a novel approach to control for spatial confounding in an outcome regression model that focuses on explaining spatial variability in the exposure.

This work is motivated by a real-world study that linked data on the built environment with electronic health record (EHR) data from an integrated healthcare system in Washington State (Mooney et al. 2020). Epidemiological studies using EHR data can be particularly subject to bias due to unmeasured spatial confounding because a key confounder of many associations of interest – SES – is not routinely available. Though EHR data studies can typically adjust for some aspects (or correlates) of SES such as race/ethnicity, insurance status, or area-level census measures, the potential for residual spatial confounding at finer scales is likely to remain. With big data EHR studies becoming increasingly common, methods to adjust for unmeasured confounding due to SES are essential to reduce the potential for these studies to yield biased health effect estimates.

The remainder of the paper proceeds as follows. In Section 2, we introduce our framework for decomposing the exposure into spatial and non-spatial components and discuss identifiability considerations. In Section 3, we describe the candidate statistical methods to adjust for spatial confounding that will be considered in our simulation study and application, including existing methods and the proposed approach. Because our motivating study is a big data application, we focus on computationally efficient spatial models that can be applied in the large data setting. Section 4 describes the simulation study methods and results. Then in Section 5, we illustrate the methods by applying them to data from approximately 120,000 participants of our motivating EHR study (Mooney et al. 2020). We consider both a purely spatial exposure (having a supermarket within walking distance of the home), as well as an individual-level exposure that clusters spatially (smoking status). Finally, in Section 6 we provide concluding remarks.

## 2. Statistical and exposure decomposition framework

We consider the setting of a person-level (point process) dataset, in which each person has been assigned a spatial coordinate (e.g., by geocoding) that captures their geographical location (e.g., residence). Let $s_i$ denote the spatial location (e.g., longitude/latitude) for person $i$ ($i = 1, ..., n$). We note that multiple individuals may have the same geocoded spatial location, for example due to multiple apartment units in the same building. In addition, we assume that $Y_i$ is the health outcome, $x_i$ is the exposure of interest that varies spatially, and $c_i$ is a vector of measured confounders (some of which may be spatially dependent). Finally, we assume that there is an unmeasured, spatially varying confounder $z_i^c$.

To allow for spatially varying exposures, we assume that the exposure of interest can be decomposed as the sum of spatially structured and "white noise" effects. Specifically, we assume

$$x_i = x^0(s_i) + \epsilon_i^x,$$

where $x^0(s_i)$ is a purely spatial component (i.e., $s_i = s_j$ implies $x^0(s_i) = x^0(s_j)$) and $\epsilon_i^x$ are independent (i.e., have no spatial correlation). To simplify notation, throughout the paper we suppress the dependence of random variables on the spatial location except for those variables that vary purely spatially (e.g., $x_i$ is not denoted as being dependent on $s_i$ but $x^0(s_i)$ is). We denote the variance of the spatial and non-spatial components as $V_0$ and $V_\epsilon$, respectively, and we define $P_{ns} = V_\epsilon/\{V_0 + V_\epsilon\}$ to be the proportion of the total variability in the exposure that can be explained by non-spatial variation. We note that this framework includes as a special case the setting of purely spatially varying exposures ($P_{NS} = 0$) considered by prior studies. We further note that although some purely spatial exposures may be recorded at an individual level (e.g., point-level predictions from a spatial model), we will refer to these as "spatial-level exposures" and will refer to exposures with a non-spatial component of variability as "individual-level exposures."

The spatially structured and white noise components of the exposure may each be subject to confounding at different levels. Confounding of the association between $x^0(s_i)$ and the outcome is possible only if there is a spatially dependent confounder (either spatial-level or person-level). On the other hand, $\epsilon_i^x$ does not vary spatially and therefore its association with the health outcome is not subject to spatial confounding (though it can clearly be subject to confounding by non-spatial variability).

For our analytic framework, we consider fitting spatial regression models taking the form (for continuous outcomes) of

$$Y_i = \alpha + \beta x_i + \gamma' c_i + g(s_i) + \epsilon_i^y, \quad (1)$$

where the parameter $\beta$ is the target of inference, $\gamma' c_i$ adjusts for measured confounders (possibly nonlinearly), and $\epsilon_i^y \sim N(0, \sigma_y^2)$ is the residual error assumed to be independent across individuals. The term $g(s_i)$ represents spatially structured variability associated with the outcome that can be modeled in different ways (see Section 3). In the setting of purely spatial exposures, as discussed by Paciorek (2010), if $g(s_i)$ is unconstrained, then $\beta x_i$ and $g(s_i)$ are not identifiable, since one could define $g^*(s_i) \equiv$

$\beta x^0(s_i) + g(s_i) = \beta x_i + g(s_i)$ without changing the likelihood. In contrast, if there is some non-spatial variability in the exposure, then $\beta$ is identifiable without constraints; in this case we have

$$Y_i = \alpha + \beta \epsilon_i^x + \boldsymbol{\gamma}' \boldsymbol{c}_i + g^*(s_i) + \epsilon_i^y$$

with $g^*(s_i)$ defined above. This suggests that if one were to control for the spatial variability in $x_i$ such that $x^0(s_i)$ gets incorporated into the spatially structured term $g^*$, this would then enable estimating the exposure-outcome association of the non-spatial component of exposure ($\epsilon_i^x$). At the end of the next section we propose a new approach to select the spatial smoothing parameter for $g$ to achieve this goal.

## 3. Statistical methods to adjust for spatial confounding in big data applications

In this section, we describe a set of candidate statistical methods that may be applied in the epidemiological literature to adjust for spatial confounding. We consider outcome regression-based methods taking the form of equation (1) to estimate the association between the exposure of interest $x_i$ and the health outcome $Y_i$, with a focus on computationally efficient approaches that can be applied to large datasets as in our application. We consider two general classes of spatial models for the spatially structured term $g(s_i)$: models that include a spatially correlated random effect and models that flexibly adjust for fixed-effect terms of spatial location using spline basis functions. In our exposition, we focus on continuous outcomes; for other outcome types (e.g., binary, count) one can apply the appropriate distribution and link function through the generalized version of the model (e.g., generalized linear mixed model [GLMM] or generalized additive model [GAM]).

### 3.1 Spatial random-effect models

Under a spatial GLMM, the vector $\boldsymbol{g}(\boldsymbol{s}) = (g(s_1), \dots, g(s_n))$ is modeled as a spatial random effect (Diggle, Tawn, and Moyeed 1998; Banerjee, Carlin, and Gelfand 2004). This random effect is commonly assumed to follow a Gaussian process with mean zero and a spatial variance-covariance matrix $V(\boldsymbol{\theta})$ having $(i,j)$ element $v_{ij} = V(s_i, s_j \mid \boldsymbol{\theta})$, where $V(\cdot,\cdot \mid \boldsymbol{\theta})$ is a parametric spatial covariance function. Inference for these models is often conducted in a Bayesian framework, with model fitting typically proceeding using a Markov chain Monte Carlo (MCMC) algorithm. Although conceptually straightforward, these models can exhibit computational difficulties, since model fitting requires computing the determinant and inverse of an n-dimensional matrix at each iteration of the MCMC algorithm.

A range of methods have been proposed to adapt the Gaussian process model to big spatial applications (Heaton et al. 2019). We considered one of these approaches, the *nearest neighbor Gaussian process (NNGP)* model, which has been found to well approximate the underlying Gaussian process (Datta et al. 2016). In addition, it achieved good predictive ability and accurate uncertainty quantification on both simulated and real data in a case study competition (Heaton et al. 2019). Briefly, the NNGP approach begins with an underlying Gaussian process for the spatial random effect $g(s)$. Given the original covariance function $V(s_i, s_j \mid \boldsymbol{\theta})$, the NNGP covariance function $\widetilde{V}(s_i, s_j \mid \boldsymbol{\theta})$ is constructed in a way that ensures that the inverse matrix $\widetilde{V}(\boldsymbol{\theta})^{-1}$ is sparse, thereby enabling the multivariate Gaussian likelihood of $g(s)$ to be evaluated efficiently. Details on the construction of the NNGP prior are given by Datta et al. (2016) and details on the software implementation are described by Finley, Datta, and Banerjee (2020).

*A note on restricted spatial regression.* An alternate spatial modeling approach that avoids the potential for increased bias due to including a spatial random effect is restricted spatial regression (RSR; Hodges and Reich 2010; Reich, Hodges, and Zadnik 2006; Hanks et al. 2015; Hughes and Haran 2013). This approach constrains/restricts the spatial random effect to the orthogonal complement of the fixed effect, such that all of the variation in the outcome over which the fixed effect and the random effect are competing is attributed to the fixed effect. Under RSR, the posterior mean of the fixed effect parameters, conditional on the variance components, is the same as under the corresponding model without the spatial random effect, while the posterior variance is changed with the goal of appropriately accounting for spatial dependence. Recent work, however, found that this is generally not the case (Khan and Calder 2019; Hanks et al. 2015). Because RSR aims to estimate the same fixed effects as the non-spatial model, in the presence of unmeasured spatial confounding, the fixed-effect estimate will have a similar bias as the non-spatial model. Since the primary goal of this work is to evaluate spatial modeling approaches based on their ability to reduce spatial confounding bias relative to a non-spatial model, we therefore do not focus on RSR methods in our simulation study or application.

### 3.2 Spatial spline models

As an alternative to the spatial GLMM approach that models spatial variability as a random effect, one can directly model the spatially structured term $g(\cdot)$ in equation (1) as a smooth function of the spatial location (Paciorek 2010). A popular choice to model $g(\cdot)$ is through spline – or piecewise polynomial – basis functions (Ruppert, Wand, and Carroll 2003). We refer to these sets of spatial spline models as spatial GAMs.

The degree of spatial smoothing (often parameterized by the degrees of freedom [DF]) can be either specified *a priori* or selected using a data-driven approach. Therefore, we considered these approaches: *fixed DF models* in which the user specifies the number of DF of the smoother and two data-driven penalized spline (PS) approaches. In the standard PS approach the DF are selected to best predict spatial variability in the outcome. In a proposed, alternate, *exposure-penalized spline* (E-PS) approach the DF are selected to best predict spatial variability in the exposure. For each of the spline-based approaches, we apply thin plate regression splines, which are computationally efficient for large datasets and do not require the specification of knot locations (Wood 2003). We fit the models using commonly available software for GAMs (mgcv R package) and apply frequentist inference (Wood 2011).

*Fixed DF models.* In these models, a fixed number of DF to be adjusted for by the spline terms is specified. We use the notation $g(s_i; K)$, where $K$ is the dimension of the basis used to represent the smooth function (the number of DF is given by $K$-1, with 1 DF lost to the identifiability constraint on the smooth). Because different values of $K$ adjust for spatial variability at different spatial scales, fixing the DF has the potential to inadequately adjust for spatial confounding (if the confounding occurs at a finer scale than the selected smoothness of $g$); consequently, this approach could lead to residual confounding bias in estimating the target parameter $\beta$. Conversely, such an approach could overadjust for spatial confounding (if the confounding occurs at a larger scale than the selected smoothness of $g$), leading to reduced efficiency in the estimate of $\beta$.

*Penalized spline models.* We also consider PS models in which the dimension of the spline basis $K$ is specified (providing an upper limit on the number of DF adjusted for), and then a penalized likelihood is employed to smooth the spline function (Wood 2017). The penalized likelihood depends on a smoothing parameter ($\lambda$) that we select using the generalized cross-validation (GCV) criterion. We use the notation $g(s_i; \lambda, K)$ to indicate that the smooth function depends both on the value of the smoothing parameter $\lambda$ and on basis dimension $K$. In principle, if $K$ is large enough that there are sufficient DF to represent the underlying spatial surface, then the estimated smooth function does not depend on the choice of $K$ (Wood 2017). Nonetheless, this should be checked for any given application. We note a correspondence between the spatial random-effect model as described in Section 3.1 and PS models (Kimeldorf and Wahba 1970); the smoothing parameter in the PS likelihood plays a role similar to the parameters $\boldsymbol{\theta}$ in the spatial covariance function.

*Exposure-penalized spline models.* Finally, motivated by the exposure decomposition framework described in Section 2, we propose an alternative approach to selecting the degree of spatial smoothing that targets confounding adjustment. Specifically, we propose a two-stage modeling approach in which the degree of smoothness is selected based on the ability to explain spatial variability in the exposure. In the first stage, we model the exposure by fitting an exposure PS model

$$x_i = \alpha_x + \boldsymbol{\gamma}'_x \boldsymbol{c}_i + g_x(s_i; \lambda_x, K) + \epsilon_i^x \qquad (2)$$

Then, using the estimated smoothing parameter $\hat{\lambda}_x$ from this model (obtained using the GCV criterion), at the second stage we fit an outcome regression model that uses the fixed smoothing parameter value obtained at the first stage,

$$Y_i = \alpha + \beta x_i + \boldsymbol{\gamma}' \boldsymbol{c}_i + g(s_i; \hat{\lambda}_x, K) + \epsilon_i^y.$$

As discussed in Section 2, by removing spatial variability in the outcome at the scale most strongly related to the exposure, this approach estimates the association between the residual non-spatial variability in the exposure and the outcome. The goal is for the residual variability in the exposure to comprise predominantly unconfounded variability.

This proposed E-PS method addresses the potential that the spatial scale most associated with variability in the outcome (as selected by the standard PS approach), may not fully adjust for spatial variability at the scale that is most confounded. Such incomplete confounding control has been found for variable selection methods that target outcome model fitting, relative to methods that directly target confounding adjustment in the non-spatial setting (Wang, Parmigiani, and Dominici 2012). In the setting of spatially varying exposures, distance-adjusted propensity scores have been proposed (Papadogeorgou, Choirat, and Zigler 2019), along with approaches to selecting the degree of spatial smoothing when the exposure is purely spatially varying (Keller and Szpiro 2020). Yet to our knowledge, approaches to select the spatial smoothing parameter to target spatial confounding adjustment in the common setting of individual-level exposures (that contain both spatial and non-spatial variability) have not been published. For such exposures, E-PS controls for spatial variability associated with the exposure, which enables estimating the exposure-outcome association of the non-spatial component of exposure; this association does not vary spatially, so is not subject to spatial confounding.

## 4. Simulation study

We conducted a simulation study to (1) explore whether (and in which settings) fitting a spatial model reduces bias and mean squared error (MSE) relative to a non-spatial model in the presence of unmeasured spatial confounding; and (2) examine which computationally efficient approach to spatial modeling for large data studies performs best in parameter estimation across a range of data-generating scenarios. We first introduce our simulation setup, including the different scenarios of data generation, details on the comparator approaches, and the criteria for evaluating the candidate methods, followed by a discussion of the simulation results. Code used to generate and fit the models is available at https://github.com/jenfb/spatial_confounding_methods.

### 4.1 Setup

*Data generation*

To generate spatially confounded data, we built upon the approaches of Paciorek (2010) and Thaden and Kneib (2018). We assumed exposure and outcome generating models as follows

$$Y_i = \beta x_i + \gamma z_i^c(s_i) + \epsilon_i^y, \quad i = 1, \ldots, n$$

$$x_i = \delta_u z_i^u(s_i) + \delta_c z_i^c(s_i) + \epsilon_i^x$$

where $x_i$ is the exposure of interest for person $i$, $z_i^c(s_i)$ is a purely spatially varying confounder (e.g., spatial-level SES), $z_i^u(s_i)$ is a spatially varying component of $x$ that is unconfounded (i.e., not associated with the outcome $Y_i$), and $\epsilon_i^x \sim N(0, \sigma_x^2)$ and $\epsilon_i^y \sim N(0, \sigma_y^2)$ are independent, non-spatial components of variability. Following Datta et al. (2016), we considered $n = 2{,}500$ observations for each dataset. We assumed the parameter values $\beta = 3$, $\gamma = 1$, and $\sigma_y^2 = 9$, based on values from Thaden and Kneib (2018), but with a larger residual variance given the larger sample size in our simulation; we also assumed that both the confounded and unconfounded spatial components of the exposure $x_i$ contributed equally ($\delta_u = \delta_c = 0.5$). In settings with no independent component of $x_i$ (that is, $z_i^u(s_i) \equiv 0$ for all $i$) or with no confounded component of $x_i$ (that is, $z_i^c(s_i) \equiv 0$ for all $i$), we instead set $\delta_u = \delta_c = \sqrt{0.5}$ to ensure that the total variability of the exposure would be the same across all scenarios.

We then considered three different values for variance of the non-spatial component of the exposure: $\sigma_x^2 \in \{0, 0.056, 0.5\}$. When $\sigma_x^2 = 0$, there is no non-spatial component of the exposure (i.e., $P_{NS} = 0$, following our notation from Section 2); this corresponds to settings in which the exposure data varies only spatially such that individuals within the same geographical location have identical exposure. Examples of these types of exposure data include measures of the built environment (e.g., residential density within an 800-meter buffer of the home location) and ambient air pollution levels. The values of $\sigma_x^2 = 0.056$ and $\sigma_x^2 = 0.5$ were selected to correspond to having 10% and 50%, respectively, of the total variability in the exposure explained by non-spatial variation ($P_{NS} = 0.1, 0.5$). Non-zero values of $\sigma_x^2$ correspond to scenarios of individual-level exposures or risk factors that may also cluster spatially (e.g., smoking status, blood pressure).

To generate spatially correlated variables, we first generated spatial locations $s_i$ by taking a sample of uniformly distributed random locations on the unit square. We then generated a vector of spatially correlated random variables $\tilde{\boldsymbol{z}}^a(\boldsymbol{s}) = (\tilde{z}_1^a(s_1), \ldots, \tilde{z}_n^a(s_n))'$ for $a \in \{c, u\}$ from independent, mean zero Gaussian processes with a Matérn variance-covariance matrices $V^a$. Specifically, we assumed that the $(i, j)$ element of $V^a$ was given by

$$\frac{2^{1-\nu}}{\Gamma(\nu)} \left(\sqrt{2\nu} \frac{d_{ij}}{\phi^a}\right)^\nu K_\nu\left(\sqrt{2\nu} \frac{d_{ij}}{\phi^a}\right),$$

where $d_{ij}$ is the Euclidean distance between the points $s_i$ and $s_j$; $\nu$ is the Matérn smoothness parameter (we set $\nu=3/2$); $\phi^a$ is the spatial decay parameter (values specified below); $\Gamma(\cdot)$ is the gamma function; and $K_\nu(\cdot)$ is the modified Bessel function of the second kind. We then took $z_i^a(s_i)$ to be the z-scored value of $\tilde{z}_i^a(s_i)$ obtained by subtracting the mean and dividing by the standard deviation of the generated vector $\tilde{\boldsymbol{z}}^a(\boldsymbol{s})$. This last step ensures that the observed variability of the simulated spatial processes is the same across different values of the spatial range parameters $\phi^c, \phi^u$ considered, since the sample variance of the simulated spatial process values in general decreases as the spatial range increases (see discussion in Paciorek (2010)). We considered all combinations of the values $\phi^u \in \{0.04, 0.15, 0.6, NA\}$ and $\phi^c \in \{0.04, 0.15, 0.6, NA\}$, where the notation $\phi^u = NA$ references the setting with no independent spatial component (that is, $z_i^u = 0$ for all $i$), and similarly $\phi^c = NA$ references the setting with no spatial confounding (that is, $z_i^c = 0$). Note that these scenarios of $(\nu, \phi^a)$ correspond to an effective range (defined as the distance at which the correlation equals 0.05) of approximately 0.11, 0.41, and 1.64, selected to yield spatial maps with fine, moderate, and large-scale spatial correlation (see **Supplemental Figure A**).

To summarize, the scenarios described above include 16 different combinations of the data-generating process for the spatial components (parameterized by $\{\phi^u, \phi^c\}$), and 3 different values of the non-spatial component of the exposure variability (parameterized by $\sigma_x^2$). For the scenarios with no independent spatial component (i.e., $\phi^u = NA$) we note that if $\sigma_x^2 = 0$, then there is no variability in $x$ that is not confounded (i.e., $x$ is completely colinear with the confounder $\tilde{z}_i^c$); thus, we considered only the two values of non-zero $\sigma_x^2$ in these scenarios. This resulted in 44 distinct data-generating scenarios (**Table**).

*Models*

To each simulated dataset, we fit a sequence of models (see **Table**). All of the models were fitted without the confounder $z_i^c$ (i.e., assuming that it was unmeasured), and for all models the inferential target was the parameter $\beta$ characterizing the effect of the exposure on the outcome.

First, we considered a non-spatial model that consisted of a standard linear regression model of the outcome $Y_i$ on the exposure variable $x_i$. We then fit a NNGP model based on an underlying exponential Gaussian process and using 10 nearest neighbors, beyond which Datta et al. (2016) found little improvement in predictive accuracy and credible interval width. To complete the model specification, we assumed flat prior distributions for the model coefficients (intercept and slope parameters), inverse Gamma priors IG(0.01, 0.01) for the variance components (residual variance and variance of the spatial random effects $g(s_i)$), and a uniform prior U(0.1, 30) for the spatial decay parameter $\kappa = 1/\phi$, which corresponds to an effective range as defined above between 0.1 and 30. We ran the MCMC algorithm for 10,000 iterations, keeping the latter half for inference. We report the parameter estimate for $\beta$ as the posterior mean and uncertainty in the estimate as the posterior standard deviation (SD).

For the fixed DF GAM methods, we considered the following choices for the number of DF: 5, 10, 25, 50, 100, 250, 500, and 1000. For the PS and E-PS approaches, we focused on versions of PS and E-PS with a large number of basis functions (basis dimension $K$ of 500 and 1000) to enable the models to fit 'wigglier' spatial surfaces, which allow for estimating fine-scale spatial variability.

*Operating characteristics evaluated*

To compare the performance of the different methods in estimating $\beta$, we calculated the bias and root MSE (rMSE) over 100 simulation repetitions. We further examined the degree to which the estimated uncertainty characterized the true uncertainty in the estimated $\beta$ by comparing the mean of the posterior SD (for NNGP) or the estimated standard error (SE) to the empirical standard error of the $\hat{\beta}$. Finally, we provide summary statistics describing the computation time of the different methods. The simulation was run across 8 cores of a virtualized Windows workstation with an Intel(R) Xeon(R) 2.6 GHz processor and 32 GB of memory.

## 4.2 Results

*Performance of the parameter estimates (bias and rMSE)*

To illustrate the relative performance of the methods, we first briefly highlight the scenario with purely spatially varying exposures ($\sigma_x^2 = 0$) that was the focus of prior work (Paciorek 2010; Keller and Szpiro 2020). **Supplemental Figure B** shows the mean estimate of $\beta$ (along with the 25th and 75th percentile estimates) across the different spatial scales of the unconfounded and confounded variability in the exposure $x$ (denoted by $\phi^u$ and $\phi^c$, respectively) in this scenario. Based on the findings of Paciorek (2010), spatial models have the potential to lead to increased bias relative to non-spatial models when $\phi^u > \phi^c$. Indeed, this finding is borne out in our results. For pairs ($\phi^u$, $\phi^c$) equal to (0.15, 0.04), (0.6, 0.04), and (0.6, 0.15) we observe that all spatial models are more biased than the non-spatial model. In contrast, when $\phi^u < \phi^c$, all of the spatial models are less biased than the non-spatial model. When $\phi^u = \phi^c$, the bias is similar across all of the methods, at the cost of considerably more variability in parameter estimates for some of the spatial models (E-PS and the fixed DF methods with larger DF).

When the exposure has some non-spatial variability ($P_{NS} > 0$), spatial models are minimally equally biased, and in some cases are substantially less biased than the non-spatial model when $\phi^u \leq \phi^c$ (results for $\sigma_x^2 = 0.056$ and $\sigma_x^2 = 0.5$ are in **Figures 1** and **2**, respectively). For example, when the confounded component has a large spatial scale ($\phi^c = 0.6$), both the fixed DF methods with a high DF and the E-PS method almost completely eliminated the confounding bias (bottom row of **Figures 1** and **2**). The PS approach is also substantially less biased than the non-spatial model but still has some residual bias relative to E-PS. Estimates from NNGP do not provide as much confounding adjustment in these scenarios; they are the closest to the non-spatial model estimates relative to the spline-based approaches to determining the degree of spatial smoothing. Even in the setting where spatial modeling has the potential to increase bias relative to the non-spatial model ($\phi^u > \phi^c$), the E-PS approach ranged from being slightly less biased ($\phi^c = 0.04$ under scenario $\sigma_x^2 = 0.056$) to substantially less biased ($\phi^c = 0.15$ under scenarios $\sigma_x^2 = 0.056$ and $\sigma_x^2 = 0.5$). Although PS was more biased than the non-spatial model in the setting of low non-spatial variability in the exposure ($\sigma_x^2 = 0.056$), it was no more biased than the non-spatial model when there was moderate non-spatial variability in the exposure ($\sigma_x^2 = 0.5$). In these scenarios with non-spatial variability in the exposure, we see that with increasing adjustment for spatial DF in the fixed DF models, bias continues to decrease and is nearly to completely eliminated across scenarios. However, the reduction in bias comes at the cost of increased variability in the parameter estimates. Overall, E-PS was the least biased of the data-driven methods for selecting the spatial DF across all scenarios with unmeasured spatial confounding.

In terms of rMSE (**Figure 3**), the optimal degree of smoothing depends on the spatial scale of the unmeasured confounder, which is around 50 DF when $\phi^c = 0.6$, around 250 DF when $\phi^c = 0.15$, and more than 1000 DF when $\phi^c = 0.04$. For the scenarios of unmeasured confounding with some non-spatial variability ($\sigma_x^2 > 0$), among the data-driven approaches to determine the degree of spatial smoothing (PS, E-PS, and NNGP), the E-PS approach achieved rMSE closest to the optimal. In contrast, in the absence of any spatial confounding, then E-PS had the greatest variability (and consequently the largest rMSE), due to adjusting for variation associated with the exposure but not the outcome (i.e., adjusting for an instrumental variable). Compared to the non-spatial model, in the presence of some non-spatial variability ($\sigma_x^2 > 0$), the E-PS approach always had lower rMSE when there was truly an unmeasured spatial confounder (**Supplemental Table B**), as did the PS and NNGP approaches with moderate non-spatial variability ($\sigma_x^2 = 0.5$).

The choice of basis dimension (K=1000 vs 500) did not have a meaningful impact on bias or rMSE for PS in any scenario or for E-PS when $\sigma_x^2 > 0$. For E-PS with $\sigma_x^2 = 0$, having a larger basis dimension may have resulted in slightly less bias but greater variability, resulting in an overall increase in rMSE (see **Supplemental Figure D**).

*Uncertainty quantification* (**Supplemental Figure C**). We observed that the estimated SE from the non-spatial model generally was either similar to or smaller than its empirical SE across values of $\sigma_x^2$. The PS method tended to underestimate the variability, with estimated SE smaller than the empirical SE. In contrast, E-PS overestimated the variability, with estimated SEs that were too large with $\sigma_x^2 = 0$, but that moved closer to the observed variability with increasing non-spatial variability. The NNGP method also generally overestimated the variability, across values of $\sigma_x^2$. The fixed DF methods underestimated the variability under low spatial DF and achieved close to the observed variability using over 100 DF.

*Computation time.* **Supplemental Table A** shows summary statistics of the estimated running time across all scenarios and all simulation repetitions for the longest running methods (the fixed DF methods with 250 or fewer DF all took less than 10 seconds). All models had a median run time of less than 6 minutes. For the spline methods, a doubling of the number of knots from 500 to 1000 resulted in a 3- to 10-fold increase in the median running time. Although the NNGP approach had the longest median run time, the method has the potential to be fit across multiple cores (which we did not do for this simulation), leading to faster running times.

*Summary of main simulation results.* Our results demonstrate that when an exposure has no non-spatial variability (an assumption that can be empirically examined, as we will illustrate below), spatial methods have the potential to either increase or decrease bias due to unmeasured spatial confounding relative to the non-spatial model (**Supplemental Figure B**) with consequential effects on rMSE (**Supplemental Table B**). Given uncertainty as to whether changes in the estimated effect are toward or away from the truth in any given application, fitting a spatial model may not be desirable to address spatial confounding. The exception is if the investigator has strong prior knowledge that the confounded source of variability in the exposure is at a larger spatial scale than the unconfounded variability. On the other hand, our results highlight that when an exposure has even a small non-spatial component of variability, fitting a spatial model (to remove confounded spatial variability) has the potential to substantially reduce bias (**Figures 1** and **2**) and rMSE in the estimated target parameter (**Figure 3**; **Supplemental Table B**). In Section 5, we will examine both settings of exposure data in our case study application.

*Additional scenarios and comparisons.* To further explore the performance of the proposed E-PS approach, we conducted additional simulation evaluations. First, we explored a more general setting that may also have an independent, unmeasured, spatially dependent predictor of the outcome (with spatial scale $\phi^y$). This setting corresponds to having unexplained spatial variation in the outcome that is not confounded. Focusing on the setting with moderate non-spatial variability ($\sigma_x^2 = 0.5$), we found that E-PS was able to perform well in alleviating confounding bias; it tended to have the lowest bias/rMSE, except in the scenario with $\phi^y < \phi^c < \phi^u$, where the PS approach performed best. Details are in **Supplemental Appendix A**.

Second, we compared a recently proposed spline-based statistical approach to account for unmeasured spatial confounding, referred to as "Spatial+" (Dupont, Wood, and Augustin 2020). As in E-PS, this method first fits the exposure model in equation (2). Then, the residuals $r_i^x$ from the exposure model replace the exposure of interest in the outcome penalized spline model, $Y_i = \alpha + \beta\, r_i^x + g(s_i; \lambda^+, K) + \epsilon_i^y$, where $g(s_i; \lambda^+, K)$ is a thin plate regression spline basis with smoothing parameter $\lambda^+$. Conceptually, these methods differ in that the degree of spatial smoothing is determined solely by the exposure model for E-PS, whereas Spatial+ also uses outcome information to estimate the spatial DF. Focusing on our original simulation setting with $\sigma_x^2 = 0.5$, we found that

Spatial+ had the second-lowest bias and rMSE (after E-PS) in settings where the spatial confounder was at a smaller spatial scale than the unconfounded component of variability in the exposure ($\phi^c < \phi^u$) : bias [rMSE] ranged from 0.11-0.32 [0.13-0.33] for Spatial+ versus 0.07-0.24 [0.11-0.26] for E-PS. When $\phi^c > \phi^u$, Spatial+ had the third-lowest bias and rMSE (after E-PS and PS): bias [rMSE] ranged from 0.08-0.18 [0.12-0.20] for Spatial+ versus 0.00-0.02 [0.08-0.08] for E-PS and 0.05-0.13 [0.09-0.14] for PS. The exception where Spatial+ outperformed E-PS occurred in the more general scenario where there was an independent, spatially dependent predictor of the outcome with $\phi^y < \phi^c < \phi^u$, though in this scenario PS outperformed both methods (bias [rMSE] was 0.02 [0.09] for PS, 0.04 [0.10] for Spatial+, and 0.07 [0.12] for E-PS; Complete results are in **Supplemental Appendix A**).

Finally, we examined the setting of a binary exposure as in our application. We found that E-PS using a generalized linear model with logit link for the exposure model performed well in reducing spatial confounding bias, even under slight misspecification of the exposure model (i.e., with exposure data generated under a probit regression model; see details in **Supplemental Appendix B**).

## 5. Application to data from the Moving to Health Study

To illustrate the spatial methods, we analyzed retrospective cohort data from the Moving to Health study (Mooney et al. 2020; Drewnowski et al. 2019; Buszkiewicz et al. 2021). The initial cohort included adult members from Kaiser Permanente Washington, an integrated health insurance and care delivery system in Washington State. Cohort members resided in King County, Washington (which includes the city of Seattle) and had at least one weight measurement during the study period (2005-2017). The study extracted information on member demographics, address history, diagnoses, and clinical visits. Home address data were cleaned and geocoded to the "rooftop" level. The geocoded addresses were linked to neighborhood-level measures of built environment, such as residential density and supermarket count within several pre-defined buffers (e.g., 1600 meters) of the home address. Geocoded addresses were also linked to home property values at the parcel level.

The primary aims of Moving to Health are to examine the impact of residential built environment on obesity and type 2 diabetes (Mooney et al. 2020). Both cross-sectional and longitudinal associations are being investigated for the main study; this application focused on the cross-sectional association between baseline exposure variables or risk factors with body weight at baseline. We focused on one built environment measure: a binary indicator of a supermarket within 1,600 meters (approximately 10-minute walk) of the home address ("supermarket availability"). By construction, the built environment predictors are purely spatially varying exposures within any calendar year (i.e., two individuals living at the same geocoded address will have the same built environment value).

To explore an individual-level exposure in which spatial confounding might still arise, we also examined the association of one of the individual-level covariates (smoking status: never, former, or current) with body weight. We focused on the association comparing "former" versus "current" smoking status, given extensive literature demonstrating that smoking cessation is related to weight gain. We hypothesized that former smokers would have higher body weights relative to current smokers (Filozof, Fernandez Pinilla, and Fernandez-Cruz 2004; Flegal et al. 1995; Williamson et al. 1991). Because smoking cessation is more common among people with more socioeconomic resources (Broms et al. 2004), we anticipate a naïve analysis would include spatial confounding.

From the initial adult cohort, we identified the set of weight measurements from clinical visits in which the patient was age 18-64, had a geocoded address at the time of the weight measure, and met other eligibility criteria (e.g., no cancer or bariatric surgery within the prior year). We then selected the first eligible weight measure during the study period to have a single weight measurement for each patient. Individuals with unknown race/ethnicity, a missing or invalid home property value, or no completed questionnaire reporting their smoking status were excluded.

**Supplemental Table C** shows the characteristics of the 117,865 patients analyzed by supermarket availability and smoking status. Patients with a supermarket within 1600 meters of their home were younger and had lower property values compared with patients without a nearby supermarket. Relative to current smokers, former smokers were older and more likely to be white and have higher home property values. Property values were also highly associated with body weight, with individuals in the lowest decile weighing on average 17.4 (95% confidence interval [CI]: 16.2, 18.5) pounds more than those in the highest decile. Because property values are related to both weight and supermarket availability and smoking status, a model that does not include property values could lead to biased associations. However, these measures are also correlated spatially (**Figure 4**), with southern regions of the county having higher proportions of current smokers, lower property values, and higher weights than northern regions; and with neighborhoods closer to water having higher property values but lower supermarket availability. Thus, fitting a spatial model could account for confounding by SES, even if no measure of individual-level SES (e.g., property values) is available.

We first considered the proportion of variability in the supermarket availability and smoking status (former vs. current) variables that could be explained by spatial versus non-spatial variability. As mentioned above, supermarket availability is a purely spatially varying exposure within a given calendar year: the proportion of non-spatial variability $P_{NS}$ is near zero (with little change in the measure over the approximately 12-year study period). For smoking status, since we do not have prior knowledge of the relative contribution of spatial versus non-spatial variability, we estimated $P_{NS}$ by fitting an unadjusted GAM of smoking status on spline terms of the spatial location. We allowed the data to determine the degree of smoothing via penalized splines and varied the total number of knots. The deviance explained by the spatial terms was approximately 3%. A spline basis dimension ($K$) of 500-1000 allowed for sufficient flexibility in the smooth function (the expected DF was estimated to be 165 and 180 under models with $K = 500$ and 1000, respectively). Thus, the proportion of non-spatial variability in smoking status is high ($P_{NS} > 95\%$), and our simulation study results suggest that spatial modeling has the potential to reduce bias and MSE relative to a non-spatial model for this risk factor.

We then fit a series of statistical models to estimate the association between smoking status and supermarket availability with body weight. We considered three sets of covariate adjustment: a "base" model that included sex, spline terms of age and height, and indicators for whether the patient had diabetes or a mood disorder; a model that additionally included insurance status and race/ethnicity, variables commonly available in EHR studies ("standard" model); and a model that included the above plus spline terms of property value, a higher quality measure of SES than is typically available for many EHR studies ("ideal" model). All spline terms of the covariates used natural cubic splines with a fixed number of 10 knots (located at deciles). For the purpose of this application, we posit that home property value is an important confounder, and that the "ideal" model will result in estimates closer to the true association than the "base" and "standard" models. In particular, we examined whether fitting a spatial model under the "standard" set of covariate adjustment moved estimates closer to the property-value adjusted estimate.

A subset of patients from the cohort had geocoded locations matching locations of other cohort members, potentially from participants living in the same household or apartment building. Because our data cannot distinguish individuals from the same household from different households in the same building, analyses are unable to account for this source of correlation. To address this limitation, we conducted sensitivity analyses in which a single patient was randomly selected from each duplicated geocode.

For each set of covariate adjustment, we fit a non-spatial regression model, as well as the spatial models considered in the simulation study. Specifically, we fit fixed DF models with the same number of DF as in the simulation, PS and E-PS both with $K = 1000$, and NNGP with 5 nearest neighbors. For the exposure model of the E-PS approach, we applied a logistic regression model that also included the same set of covariates as the respective outcome model. For the NNGP approach, because of repeated geocodes, to fit the model we introduced a smaller jitter (of 0.05) to duplicate geocodes to enable the model fitting. (Not doing so yielded an error). The prior distributions for the NNGP approach were the same as in the simulation study, except for the spatial decay parameter $\kappa = 1/\phi$, which was assumed to be U(0.02, 5), corresponding to an effective range between 0.7 and 181 km. (The maximal distance between residential locations in our sample was 93 km apart).

**Figure 5** shows estimates of the associations between supermarket availability (**5A**) and smoking status (**5B**) with body weight across the fitted models, along with 95% CIs (or corresponding posterior mean estimates and credible intervals for NNGP). We also see that under the non-spatial model, the estimated weight change associated with both the supermarket and smoking status exposure variables increased in magnitude with additional adjustment across the base, standard, and ideal models. (The association for supermarket availability was -2.8, -3.1, and -4.0 pounds under the three models, respectively, and the association for smoking status was 1.4, 2.1, and 3.4 pounds). This suggests that although adjustment for individual-level SES-related factors commonly available in EHR data (i.e., insurance status and race/ethnicity) may reduce bias relative to a base model that does not include these variables, meaningful confounding still occurs due to individual-level SES (as quantified by home property values).

The impact of spatial modeling on parameter estimates differed substantially across the two exposure variables. For the supermarket measure that comprised almost entirely spatial variability, all the spatial models moved the estimates further from the association under the "ideal" non-spatial model (dashed horizontal line in **Figure 5**). Except for the NNGP approach that estimated an association very similar to the corresponding non-spatial model, the other spatial modeling approaches led to estimated associations either in the opposite direction as the estimates under the non-spatial model (fixed DF methods with DF ≤250 and PS) or close to the null with large CIs (fixed DF methods with high DF and E-PS). Prior literature on supermarket availability and obesity found mostly null or negative associations (Cobb et al. 2015), whereas spatial adjustment (except at the largest spatial scales) yielded a positive association contrary to expectation.

In contrast, for the individually varying smoking status variable, spatial modeling approaches generally led to parameter estimates that were closer to the non-spatial estimate under the "ideal" model. In addition, we note that the direction of the associations is consistent with prior literature on the relations between smoking cessation and obesity (Filozof, Fernandez Pinilla, and Fernandez-Cruz 2004; Flegal et al. 1995). This suggests that even in a study that does not have property values available, adjustment for spatial variability has the potential to substantially reduce bias. (It is also possible, though, that spatial factors other than property values may drive this change in the point estimate). Interestingly, even in the ideal model, further spatial adjustment slightly increased the point estimates, indicating possible additional unmeasured spatial confounding (e.g., from environmental factors, educational attainment, or other SES aspects not captured by property values). As in the simulation study, estimates from NNGP were between those from the non-spatial model and those from the spline methods that estimated the spatial DF (PS and E-PS); interval widths were somewhat larger. Computation time for all methods was less than 60 minutes (**Supplemental Table A**), except for E-PS for the supermarket measure, which had difficulty fitting the exposure model due to overfitting (given the nearly purely spatially varying exposure).

Sensitivity analyses indicated that (1) varying the prior distribution on the spatial scale parameter for the NNGP model had negligible impact on the coefficient estimates, and

(2) randomly selecting a single individual from each geocoded address having multiple individuals did not change our conclusions (**Supplemental Appendix C**).

## 6. Discussion

This paper demonstrates a method for epidemiological studies to potentially reduce bias from unmeasured spatial confounding. Using both a simulation study and a real data application, we showed that E-PS spatial modeling can greatly reduce or even entirely remove spatial confounding bias for individual-level factors (such as smoking status) that can vary both among and within spatial locations. Moreover, we found that the higher the proportion of non-spatial variability in the exposure (a quantity that can be estimated), the greater the potential that bias will be reduced by spatial modeling. However, for exposure variables with purely spatial variation (such as built environment exposures), spatial modeling can either increase or decrease bias relative to fitting a non-spatial model.

Our simulation studies showed that for individual-level exposures, the proposed E-PS approach achieved the best performance (measured as MSE), among the data-driven methods we examined for estimating the degree of spatial smoothing, including PS and NNGP models. In the presence of unmeasured spatial confounding, E-PS "removes" confounded spatial variation to estimate the association of the unconfounded non-spatial component of variability. In some cases, using a fixed number of DF determined a priori might perform better the E-PS approach. However in practice, the optimal DF is typically unknown, so we do not know when fixed DF performs better. Nonetheless, results from fixed DF spline models could be presented in sensitivity analyses to illustrate the sensitivity of results to spatial adjustment.

Prior literature on spatial confounding suggested that statistical analyses should either avoid standard (i.e., non-restricted) spatial modeling approaches such as spatial GLMM because they have the potential to increase bias in some settings (Hodges and Reich 2010), or should rely on a potentially strong assumption regarding the relative scale of unconfounded versus confounded spatial variability in the exposure (Paciorek 2010). Our simulation results confirm the potential risks of increasing bias and MSE when fitting a standard spatial model with exposure data that vary purely spatially, particularly for spline models with a large number of DF. Thus, we do not recommend fitting these models for data with little to no non-spatial variability. However, many studies – including many EHR data studies – consider the health effects of individually varying exposures with non-spatial variability. Examples are behavioral factors (e.g., smoking status), medication use, and clinical markers (e.g., blood pressure) that are likely to be related to SES and thus vary spatially and be subject to spatial confounding (Parnia and Siddiqi 2020; Drewnowski, Buszkiewicz, and Aggarwal 2019; McDoom et al. 2020). Such studies could benefit from spatial modeling.

Recent work has highlighted sets of assumptions that enable causal interpretation of exposure effects from spatial models (Schnell and Papadogeorgou 2020; Guan et al. 2020; Reich et al. 2020). Schnell and Papadogeorgou (2020) frame the assumption on the relative spatial scales of the confounder and exposure (which they refer to as the 'spatial scale restriction') as a positivity assumption. For purely spatial exposures, they note that if the spatial scale of the unmeasured confounder is smaller than the spatial scale of the exposure then positivity is violated, since some 'strata' of the unmeasured confounder may have only a single possible value for the exposure. In contrast, for individual-level exposures, as long as a spatial location can have individuals who are both exposed and unexposed, then positivity is satisfied, allowing for identification of exposure effects. Guan et al. (2020) provide an alternate framing of the spatial scale restriction when the exposure and unmeasured spatial confounder are parameterized in the spectral domain. They show that a necessary assumption is that the 'coherence' (correlation at different spatial scales) between the exposure and confounder dissipates at higher frequencies, meaning there is no unmeasured confounding at local spatial scales. For individual-level exposures, the presence of non-spatial variability in the exposure ($\epsilon^x$) allows the identification of exposure effects even with an unmeasured spatial confounder with local (fine-scale) spatial variability. Nonetheless, achieving unbiased health effect estimation requires no unmeasured confounder of the non-spatial (individual-level) variability.

Our methods rely on the availability of spatial point data. In many epidemiologic studies, notably including cohorts assembled from data collected in routine clinical care, participant home addresses are available as text. Geocoding services, including ArcGIS, SAS macros, and open source tools (https://github.com/uwrit/postgis-docker), can be used to identify a spatial point for these addresses without compromising privacy (Bader, Mooney, and Rundle 2016). Thus, spatial models can often be fit using routinely available data.

This work has several limitations. First, although we considered several spline-based spatial modeling approaches along with a computationally efficient approximation to the Gaussian process (NNGP), other spatial models for big data have been proposed (Heaton et al. 2019). In evaluations for predictive accuracy, many of these approaches perform well. However, they have not been evaluated for their ability to reduce bias due to unmeasured confounding relative to the spline and NNGP methods considered here. Second, our simulation studies considered scenarios with only a single unmeasured spatial confounder. In practice a study could have multiple unmeasured spatial confounders at different spatial scales. A potential limitation of E-PS is that the estimates can depend on the choice of the basis dimension $K$ (though we did not find this to be the case in our simulation study). In principle, one should pick a large value, since the basis dimension corresponds to the maximal amount of spatial DF adjustment. However, increasing $K$ increases the computation time substantially (nonlinearly). We found that the computation time for E-PS was relatively fast for our application of approximately 120,000 records and for $K$ up to 1000, which was reasonable for our setting within a densely populated area. In big data applications covering a larger geographical area, or for which finer scale spatial adjustment is necessary, approaches to partitioning the data across multiple geographical regions, followed by fitting a series of regional models and then pooling resulting estimates, may be necessary.

In conclusion, based on simulations and a demonstration with real-world data, E-PS appeared promising for reducing spatial confounding bias. We propose that future work should determine its optimal implementation. For example, in the binary exposure setting, using a logistic exposure regression model performed well at reducing bias even when the true exposure-generating model was a probit regression. Further work could explore if model selection on the exposure model improves E-PS performance. We are also interested in testing the performance of the method in complex settings with multiple spatial confounders at different spatial scales.

**Supplemental Material**

Supplemental Appendixes, Figures, and Tables are available in the online supplement.

**Acknowledgments**

This work was supported by National Institute of Diabetes and Digestive and Kidney Diseases (NIDDK) Grant R01 DK114196 and by grant R00LM01286. We thank Chris Tachibana, PhD, for scientific editing.

**Data and Software Availability**

Simulated data that support the findings of this study are available on request from the corresponding author. Electronic health record data used in the application are not publicly available due to privacy or ethical restrictions. Code is available at https://github.com/jenfb/spatial_confounding_methods.

**Figure Legends**

**Figure 1.** Mean (interquartile range) of $\beta$ parameter estimates for the setting of low non-spatial variability in the exposure ($\sigma_x^2 = 0.056$; $P_{NS} = 10\%$) across different spatial scales of the unconfounded and confounded variability ($\phi^u$ and $\phi^c$, respectively). Dashed horizontal line corresponds to the true value of $\beta$; dotted horizontal line corresponds to the mean estimate from the non-spatial model. NS, non-spatial; F-DF, fixed degrees of freedom; PS, penalized spline; E-PS, exposure-penalized spline; NNGP, nearest neighbor Gaussian process.

**Figure 2.** Mean (interquartile range) of $\beta$ parameter estimates for the setting of moderate non-spatial variability in the exposure ($\sigma_x^2 = 0.5$; $P_{NS} = 50\%$) across different spatial scales of the unconfounded and confounded variability ($\phi^u$ and $\phi^c$, respectively). Dashed horizontal line corresponds to the true value of $\beta$; dotted horizontal line corresponds to the mean estimate from the non-spatial model. NS, non-spatial; F-DF, fixed degrees of freedom; PS, penalized spline; E-PS, exposure-penalized spline; NNGP, nearest neighbor Gaussian process.

**Figure 3.** Root mean squared error (rMSE) of $\beta$ parameter estimates across different values of the proportion of non-spatial variability in exposure (controlled by $\sigma_x^2$) and different spatial scales of the unconfounded and confounded variability ($\phi^u$ and $\phi^c$, respectively). Note that we truncate the y-limits of the plot, because the rMSE is too large under certain scenarios with $\sigma_x^2 = 0$ for some methods (fixed DF with high number of DF and E-PS for $\phi^u = 0.6$), such that including them masks the remaining values. NS, non-spatial; F-DF, fixed degrees of freedom; PS, penalized spline; E-PS, exposure-penalized spline; NNGP, nearest neighbor Gaussian process.

**Figure 4.** Spatial variation in variables of interest in the application to Moving to Health data. Smoothed maps for the individual-level variables (weight, smoking status, and property values) estimated from an unadjusted generalized additive model on spline terms of the spatial location. Map for the spatial-level variable (supermarket availability) shown for the year 2015.

**Figure 5.** Point estimates of the association of (A) supermarket availability and (B) smoking status (former vs. current) with body weight in the Moving to Health application, across covariate adjustment model (base, standard, and ideal). Intervals shown are 95% confidence intervals for the frequentist approaches and 95% posterior credible intervals for nearest neighbor Gaussian process (NNGP). Dashed horizontal line corresponds to the parameter estimate under the non-spatial model with "ideal" covariate adjustment. NS, non-spatial; F-DF, fixed degrees of freedom; PS, penalized spline; E-PS, exposure-penalized spline.

**Table.** Overview of simulation to compare big data methods to adjust for unmeasured spatial confounding

### Data generating scenarios

| Parameter | Definition | Values | Explanation |
|---|---|---|---|
| $\sigma_x^2$ | Variance of non-spatial error in the exposure $x$ | 0, 0.056, 0.5 | The selected values of $\sigma_x^2$ correspond to different values of the proportion $P_{NS}$ of total variability in the exposure that is explained by non-spatial variability:<br>• no non-spatial variability ($\sigma_x^2 = 0$)<br>• low (10%) non-spatial variability (0.056)<br>• moderate (50%) non-spatial variability (0.5) |
| $\phi^u$ | Spatial range of the unconfounded variability in the exposure $x$ | 0.04, 0.15, 0.6, NA | Spatial range of the component $z^u$ of $x$ that is not related with the health outcome.<br>• $\phi^u = NA$ indicates the scenario with no independent spatial component (that is, $z_i^u \equiv 0$ for all individuals $i$)<br><br>The other values correspond to spatial variability at the following scales (see **Supplemental Figure A**):<br>• Fine ($\phi^u = 0.04$)<br>• Moderate ($\phi^u = 0.15$)<br>• Large ($\phi^u = 0.6$) |
| $\phi^c$ | Spatial range of the confounded variability in the exposure $x$ | 0.04, 0.15, 0.6, NA | Spatial range of the component $z^c$ of $x$ that is related with the health outcome.<br>• $\phi^c = NA$ references the setting with no spatial confounding (that is, $z_i^c \equiv 0$ for all individuals $i$)<br><br>The other values correspond to spatial variability at the following scales (see **Supplemental Figure A**):<br>• Fine ($\phi^u = 0.04$)<br>• Moderate ($\phi^u = 0.15$)<br>• Large ($\phi^u = 0.6$) |

### Comparator spatial modeling approaches

| Name | Description | Variations considered |
|---|---|---|
| NS | Non-spatial model | |
| F-DF | Fixed degrees of freedom spline model | Degrees of freedom*: 5, 10, 25, 50, 100, 250, 500, 1000 |
| PS | Penalized spline model | Spline basis dimension (K): 500, 1000 |
| E-PS | Exposure-penalized spline model | Spline basis dimension (K): 500, 1000 |
| NNGP | Nearest neighbor Gaussian process | |

*Note: for the F-DF spline models, the number of DF is equal to K-1, where K is the dimension of the spline basis

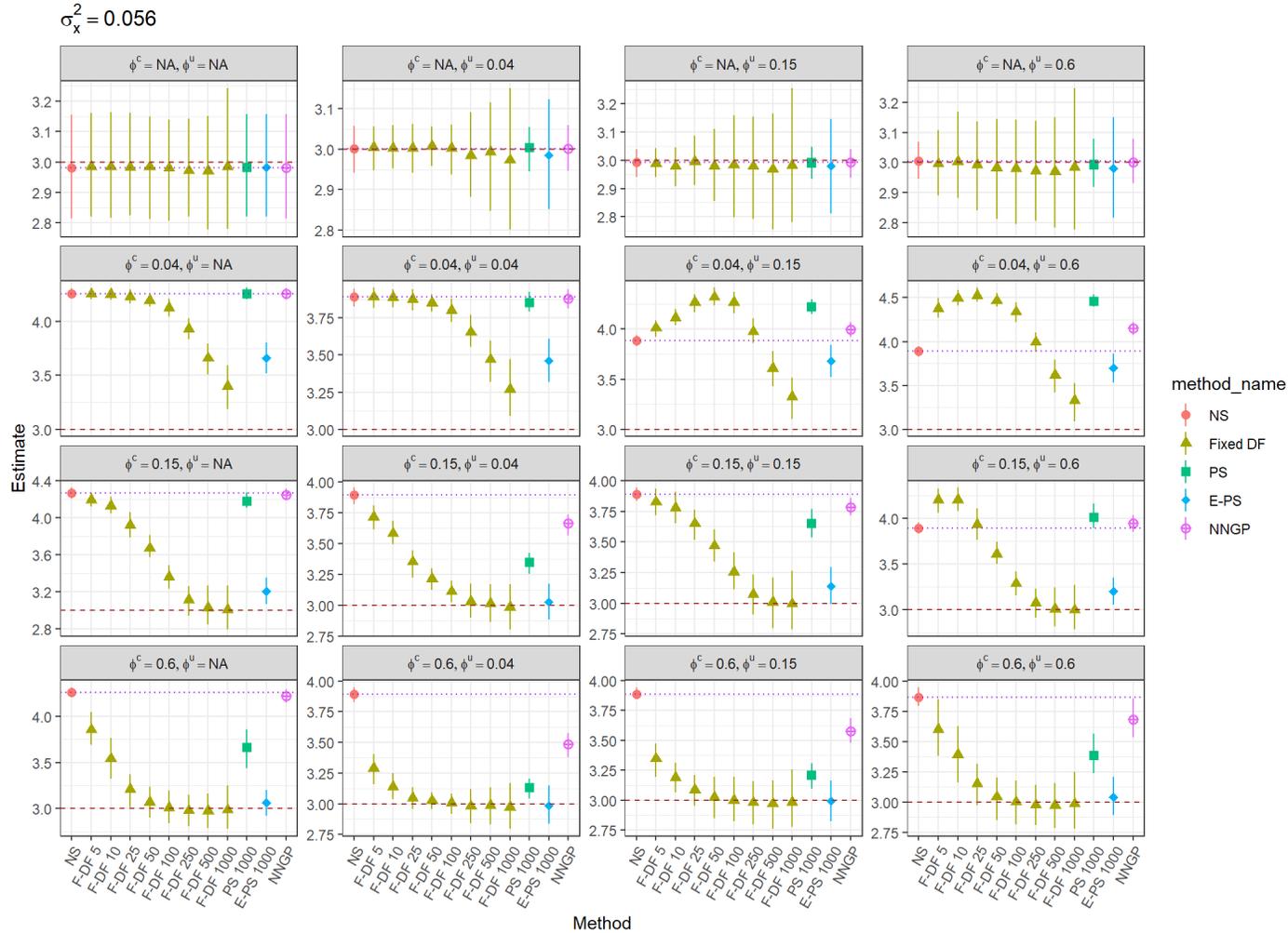

**Figure 1.** Mean (interquartile range) of $\beta$ parameter estimates for the setting of low non-spatial variability in the exposure ($\sigma_x^2 = 0.056$; $P_{NS} = 10\%$) across different spatial scales of the unconfounded and confounded variability ($\phi^u$ and $\phi^c$, respectively). Dashed horizontal line corresponds to the true value of $\beta$; dotted horizontal line corresponds to the mean estimate from the non-spatial model. NS, non-spatial; F-DF, fixed degrees of freedom; PS, penalized spline; E-PS, exposure-penalized spline; NNGP, nearest neighbor Gaussian process.

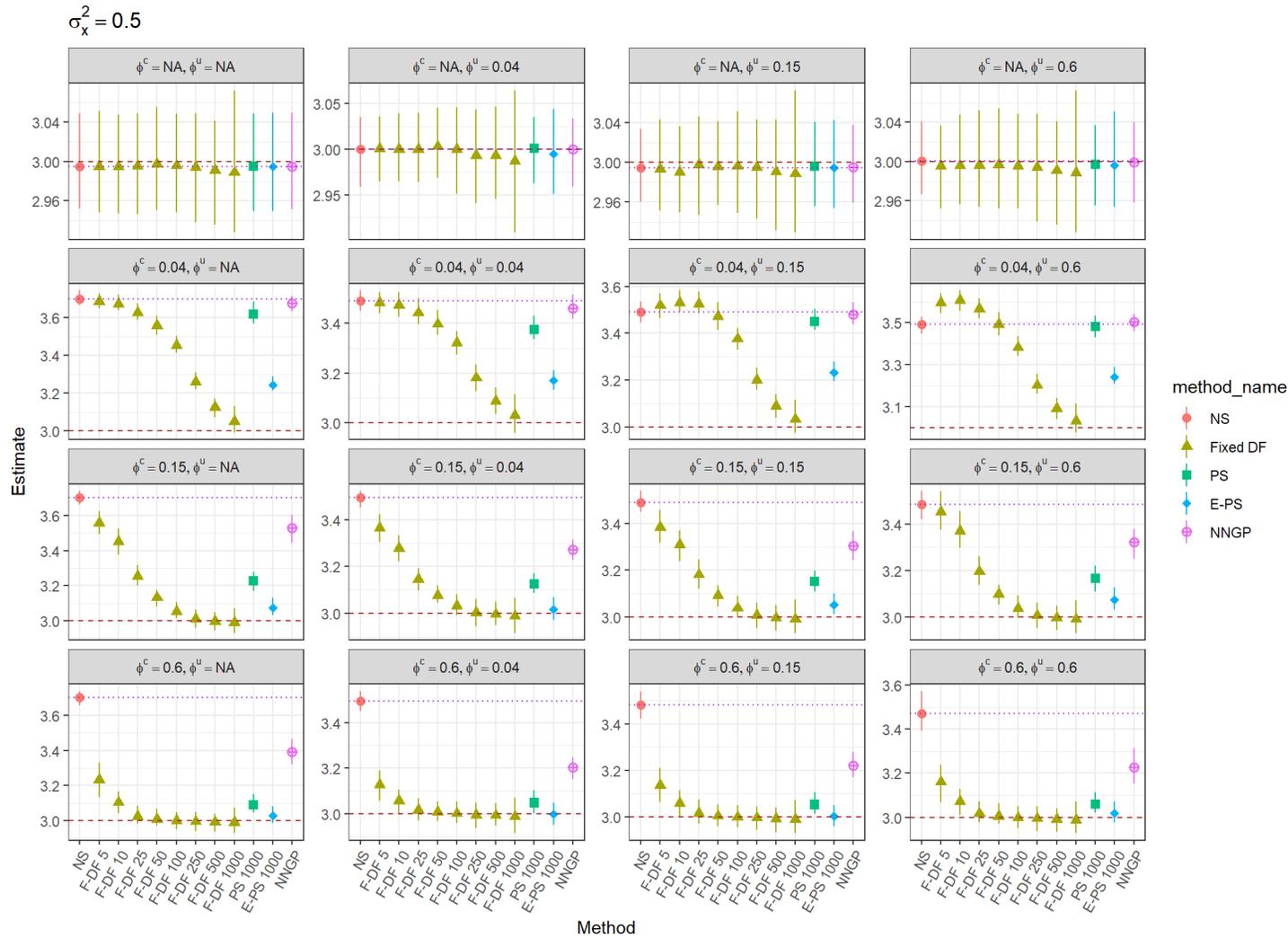

**Figure 2.** Mean (interquartile range) of $\beta$ parameter estimates for the setting of moderate non-spatial variability in the exposure ($\sigma_x^2 = 0.5$; $P_{NS} = 50\%$) across different spatial scales of the unconfounded and confounded variability ($\phi^u$ and $\phi^c$, respectively). Dashed horizontal line corresponds to the true value of $\beta$; dotted horizontal line corresponds to the mean estimate from the non-spatial model. NS, non-spatial; F-DF, fixed degrees of freedom; PS, penalized spline; E-PS, exposure-penalized spline; NNGP, nearest neighbor Gaussian process.

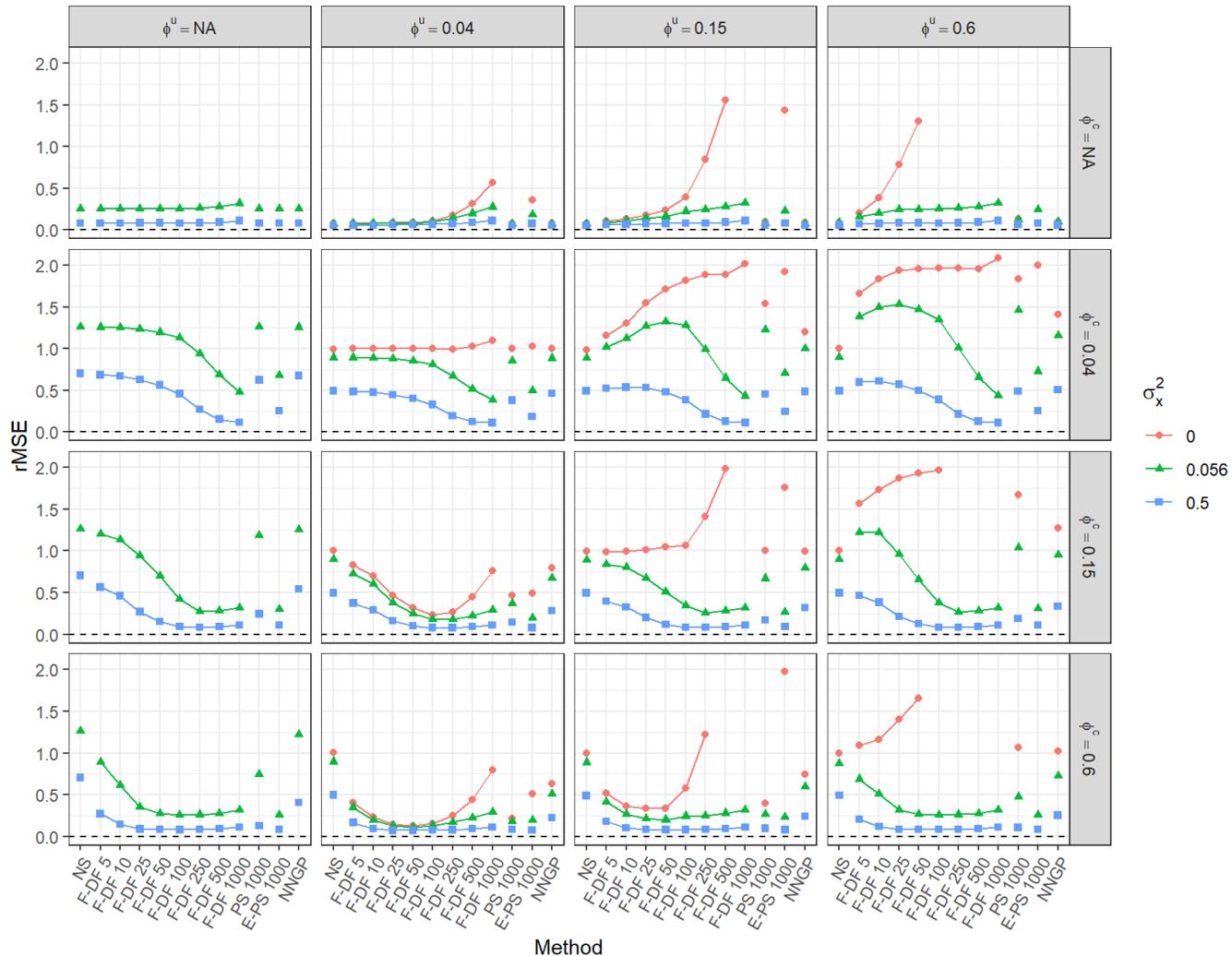

**Figure 3.** Root mean squared error (rMSE) of $\beta$ parameter estimates across different values of the proportion of non-spatial variability in exposure (controlled by $\sigma_x^2$) and different spatial scales of the unconfounded and confounded variability ($\phi^u$ and $\phi^c$, respectively). Note that we truncate the y-limits of the plot, because the rMSE is too large under certain scenarios with $\sigma_x^2 = 0$ for some methods (fixed DF with high number of DF and E-PS for $\phi^u = 0.6$), such that including them masks the remaining values. NS, non-spatial; F-DF, fixed degrees of freedom; PS, penalized spline; E-PS, exposure-penalized spline; NNGP, nearest neighbor Gaussian process.

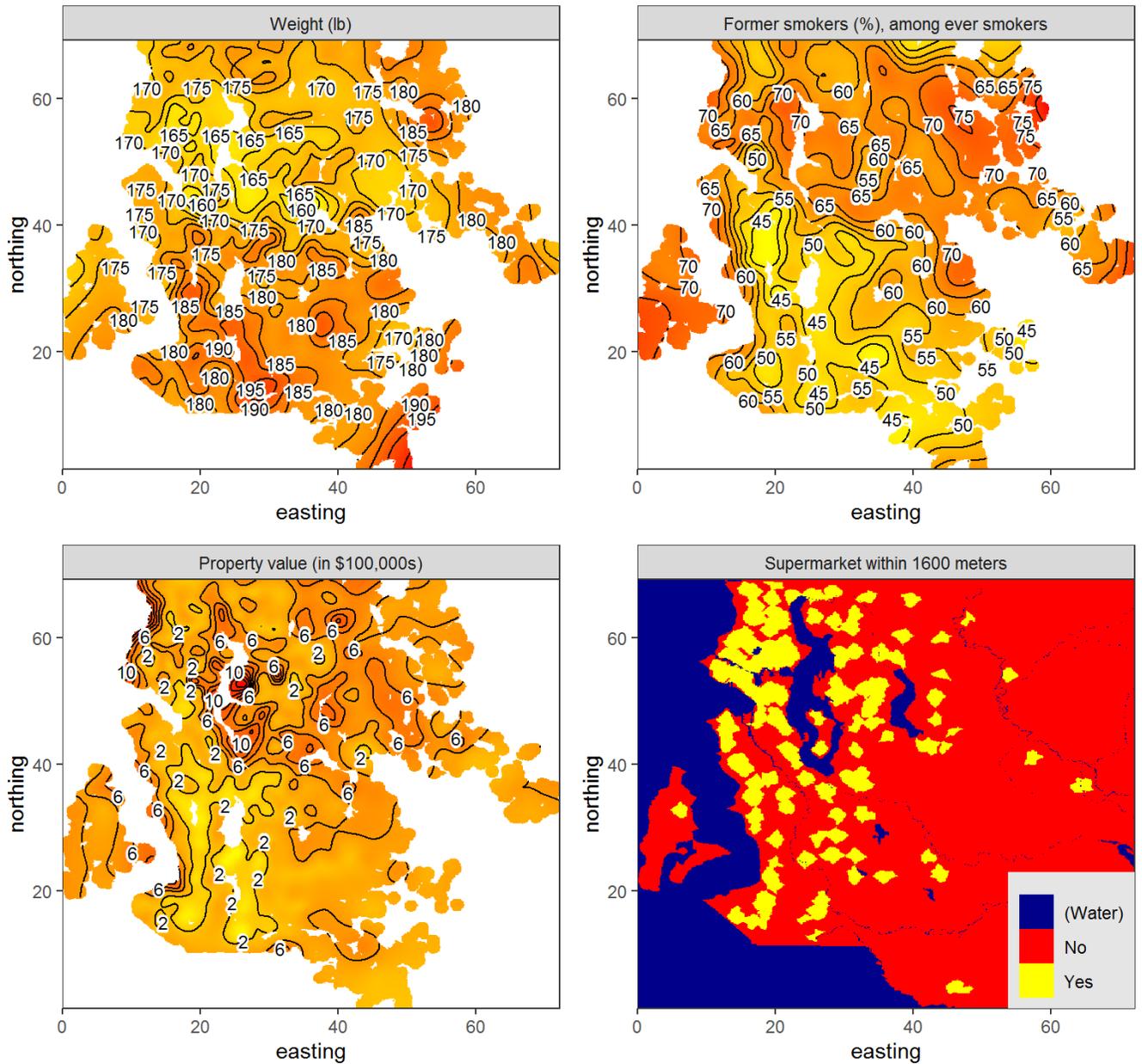

**Figure 4.** Spatial variation in variables of interest in the application to Moving to Health data. Smoothed maps for the individual-level variables (weight, smoking status, and property values) estimated from an unadjusted generalized additive model on spline terms of the spatial location. Map for the spatial-level variable (supermarket availability) shown for the year 2015.

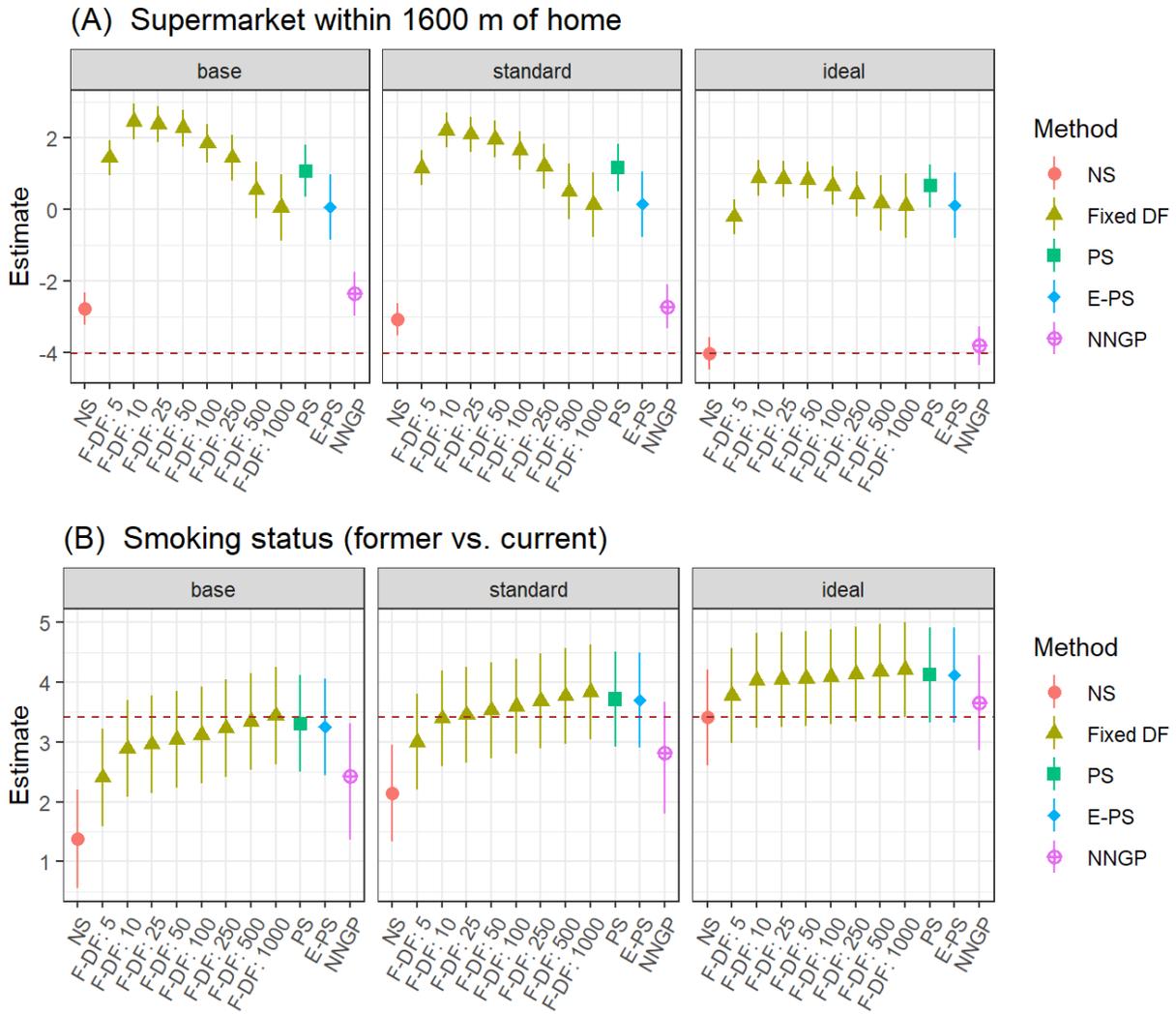

**Figure 5.** Point estimates of the association of (A) supermarket availability and (B) smoking status (former vs. current) with body weight in the Moving to Health application, across covariate adjustment model (base, standard, and ideal). Intervals shown are 95% confidence intervals for the frequentist approaches and 95% posterior credible intervals for nearest neighbor Gaussian process (NNGP). Dashed horizontal line corresponds to the parameter estimate under the non-spatial model with "ideal" covariate adjustment. NS, non-spatial; F-DF, fixed degrees of freedom; PS, penalized spline; E-PS, exposure-penalized spline.

Web-based supporting material for "Accounting for spatial confounding in epidemiological studies with individual-level exposures: an exposure-penalized spline approach"

by

Jennifer F. Bobb, Maricela F. Cruz, Stephen J. Mooney, Adam Drewnowski, David Arterburn, Andrea J. Cook

**Supplemental Appendix A.** Additional simulation scenarios that included an independent spatial component of the outcome, along with comparison of Spatial+ approach.

In our original simulation study, we assumed that spatial variation in the outcome arose due to spatial variation in the exposure along with spatial variation due to an unmeasured spatial confounder. However, it is possible that there may be additional sources of (unmeasured) spatial variation in the outcome beyond these two sources. To address this more general scenario, we expand our data generating model as follows:

$$Y_i = \beta x_i + \gamma z_i^c(s_i) + \gamma_y z_i^y(s_i) + \epsilon_i^y, \quad i = 1, \ldots, n$$
$$x_i = \delta_u z_i^u(s_i) + \delta_c z_i^c(s_i) + \epsilon_i^x$$

where the only new component beyond those defined in the main manuscript text is $z_i^y(s_i)$, an independent (i.e., unconfounded) spatial predictor of the outcome. We assume that $\gamma_y = 1$ with the other parameters as in our original simulation study. Similar to the other spatial components, we assume that $z_i^y(s_i)$ comes from a Gaussian process with Matérn covariance having smoothness parameter $\nu=3/2$ and spatial decay parameter $\phi^y \in \{0.04, 0.15, 0.6\}$. For these additional scenarios, we focus on the setting where the proportion of non-spatial variability was moderate ($\sigma_x^2 = 0.5$), and on settings where the three spatial components ($z^c, z^u, z^y$) have different spatial scales, leading to six different scenarios of ($\phi^c, \phi^u, \phi^y$).

For each simulated dataset, we fit the non-spatial model (NS), the nearest neighbor Gaussian process approach (NNGP), as well as the penalized spline (PS) and exposure-penalized spline (E-PS) methods. We additionally evaluated the newly proposed Spatial+ (Spat+) method that replaces the exposure in the PS model with residuals from the exposure model that adjusts for splines of the spatial location (Dupont et al. 2020). As in the simulation study of Dupont et al. (2020), we use the generalized cross validation (GCV) criterion for smoothing parameter estimation. All of the spline-based methods used a basis dimension of $K = 500$. We evaluated the methods in terms of bias (where the true value of $\beta$ was 3), root mean squared error (rMSE), and by the ratio of the estimated standard error (SE) divided by the empirical SE. The following table shows the results of these additional scenarios side by side with the corresponding scenario from our original simulation study in which there was no independent spatial predictor of the outcome.

**Table.** Performance of methods under additional simulation scenarios that include an independent spatial component $z^y(s)$ of the outcome as compared to the original scenarios. Bolded cells indicate the method with the smallest bias or rMSE.

| Scenario | | | Original setting ($\phi^y = NA$) | | | Spatial component of outcome | | |
|---|---|---|---|---|---|---|---|---|
| $\phi^c$ | $\phi^u$ | Method | Bias | rMSE | $\widehat{SE}(\beta)/SD(\hat{\beta})$ | Bias | rMSE | $\widehat{SE}(\beta)/SD(\hat{\beta})$ |
| $\phi^c < \phi^u$ | | | | | | | | |
| | | | | | | $\phi^y = 0.6$ | | |
| 0.04 | 0.15 | NS | 0.49 | 0.50 | 0.92 | 0.47 | 0.55 | 0.25 |
| 0.04 | 0.15 | PS | 0.45 | 0.46 | 0.88 | 0.43 | 0.44 | 0.93 |
| 0.04 | 0.15 | E-PS | **0.24** | **0.25** | 1.03 | **0.24** | **0.25** | 1.03 |
| 0.04 | 0.15 | Spat+ | 0.31 | 0.33 | 1.04 | 0.31 | 0.33 | 1.04 |
| 0.04 | 0.15 | NNGP | 0.48 | 0.49 | 1.02 | 0.42 | 0.43 | 1.35 |
| | | | | | | $\phi^y = 0.15$ | | |
| 0.04 | 0.6 | NS | 0.49 | 0.50 | 1.00 | 0.51 | 0.57 | 0.29 |
| 0.04 | 0.6 | PS | 0.48 | 0.49 | 0.79 | 0.37 | 0.38 | 0.98 |
| 0.04 | 0.6 | E-PS | **0.25** | **0.26** | 1.03 | **0.25** | **0.26** | 1.03 |
| 0.04 | 0.6 | Spat+ | 0.32 | 0.33 | 1.04 | 0.31 | 0.32 | 1.03 |
| 0.04 | 0.6 | NNGP | 0.50 | 0.51 | 1.09 | 0.38 | 0.39 | 1.46 |
| | | | | | | $\phi^y = 0.04$ | | |
| 0.15 | 0.6 | NS | 0.49 | 0.49 | 0.64 | 0.47 | 0.51 | 0.42 |
| 0.15 | 0.6 | PS | 0.17 | 0.19 | 0.98 | **0.02** | **0.09** | 1.03 |
| 0.15 | 0.6 | E-PS | **0.07** | **0.11** | 1.07 | 0.07 | 0.12 | 0.97 |
| 0.15 | 0.6 | Spat+ | 0.11 | 0.13 | 1.08 | 0.04 | 0.10 | 1.03 |
| 0.15 | 0.6 | NNGP | 0.32 | 0.34 | 1.32 | 0.15 | 0.19 | 1.65 |
| $\phi^c > \phi^u$ | | | | | | | | |
| | | | | | | $\phi^y = 0.6$ | | |
| 0.15 | 0.04 | NS | 0.49 | 0.50 | 1.04 | 0.51 | 0.58 | 0.27 |
| 0.15 | 0.04 | PS | 0.13 | 0.14 | 1.02 | 0.11 | 0.13 | 1.06 |
| 0.15 | 0.04 | E-PS | **0.02** | **0.08** | 1.08 | **0.02** | **0.08** | 1.08 |
| 0.15 | 0.04 | Spat+ | 0.18 | 0.20 | 1.04 | 0.18 | 0.20 | 1.04 |
| 0.15 | 0.04 | NNGP | 0.27 | 0.28 | 1.82 | 0.17 | 0.19 | 2.17 |
| | | | | | | $\phi^y = 0.15$ | | |
| 0.6 | 0.04 | NS | 0.49 | 0.50 | 0.98 | 0.50 | 0.57 | 0.26 |
| 0.6 | 0.04 | PS | 0.05 | 0.09 | 0.97 | 0.01 | **0.08** | 1.01 |
| 0.6 | 0.04 | E-PS | **0.00** | 0.08 | 1.05 | **0.00** | **0.08** | 1.05 |
| 0.6 | 0.04 | Spat+ | 0.18 | 0.20 | 1.02 | 0.16 | 0.18 | 1.02 |
| 0.6 | 0.04 | NNGP | 0.20 | 0.22 | 1.68 | 0.08 | 0.11 | 2.22 |
| | | | | | | $\phi^y = 0.04$ | | |
| 0.6 | 0.15 | NS | 0.48 | 0.49 | 0.76 | 0.47 | 0.50 | 0.45 |
| 0.6 | 0.15 | PS | 0.05 | 0.10 | 0.95 | -0.01 | **0.09** | 1.01 |
| 0.6 | 0.15 | E-PS | **0.00** | **0.08** | 1.04 | **0.00** | 0.09 | 1.01 |
| 0.6 | 0.15 | Spat+ | 0.08 | 0.12 | 1.04 | 0.02 | **0.09** | 1.01 |
| 0.6 | 0.15 | NNGP | 0.22 | 0.24 | 1.69 | 0.12 | 0.16 | 1.99 |

In comparing Spatial+ to the other methods under the original scenarios, we see that it had the second lowest bias/rMSE after the E-PS approach in settings where the spatial confounder was at a smaller spatial scale than the unconfounded component of variability in the exposure ($\phi^c < \phi^u$). When $\phi^c > \phi^u$ it had the third lowest bias/rMSE after E-PS and PS.

In the additional simulation scenarios that had an independent spatial component associated with the outcome, we found that E-PS tended to have the lowest bias/rMSE, except in the scenario with $\phi^y < \phi^c < \phi^u$, where the PS approach performed best, followed by Spatial+ and then E-PS. Because PS uses outcome information to identify the spatial scale that is adjusted for in the outcome model, when there is an independent spatial component $z^y$ of the outcome that varies at a finer spatial scale than the confounded variability (i.e., $\phi^y < \phi^c$), PS accounts for spatial variability at this finer spatial scale. Consequently, the PS approach is able to incidentally also reduce unmeasured confounding bias. In contrast, when the independent spatial component $z^y$ is at a larger spatial scale ($\phi^y = 0.6$) or there is no independent spatial component of the outcome (original scenarios) PS remains substantially more biased than E-PS. As in the original simulation study scenarios, E-PS tended to slightly overestimate uncertainty with estimated SEs larger than the empirical SEs, except in the scenario where $(\phi^c, \phi^u, \phi^y) = (0.15, 0.6, 0.04)$ where it slightly underestimated the variability. This is not surprising because E-PS is attempting to select the spatial scale to match $\phi^c$, which does not fully capture the finer spatial variability in the outcome.

**Supplemental Appendix B.** Additional simulation scenarios of the binary exposure setting.

To evaluate the performance of the E-PS approach in the scenario of a binary exposure as in our application, we generated exposure data as

$$x_i = I(x_i^* > 0)$$
$$x_i^* = \delta_u z_i^u(s_i) + \delta_c z_i^c(s_i) + \epsilon_i^x$$

where $\epsilon_i^x$ is standard normal and $I(\cdot)$ denotes the indicator function. We note that this exposure generating model corresponds to the latent variable formulation of a probit regression model. The other parameters were the same as in our original simulation, as was the outcome generating model for $Y_i$. We considered three values of the spatial range parameter ($\phi$) as in our original simulation $\{0.04, 0.15, 0.6\}$. We focused on settings in which $(z^c, z^u)$ each have different spatial scales, leading to six different combinations of $(\phi^c, \phi^u)$.

For each simulated dataset, we fit the non-spatial model (NS), the penalized spline (PS), and exposure-penalized spline (E-PS) methods. For E-PS, we considered three different versions, in which we varied the fitted exposure model as (1) a generalized linear model (GLM) with logit link, (2) a GLM with probit link, or (3) a linear regression model. All of the spline-based methods used a basis dimension of $K = 500$. We evaluated the methods in terms of bias (where the true value of $\beta$ was 3), root mean squared error (rMSE), and by the ratio of the estimated standard error (SE) divided by the empirical SE. The following table shows the results of these additional scenarios.

**Table.** Performance of methods under setting of binary spatial exposure. Bolded cells indicate the method with the smallest bias or rMSE.

| Scenario | | | Performance Metric | | |
|---|---|---|---|---|---|
| $\phi^c$ | $\phi^u$ | Method | Bias | rMSE | $\widehat{SE}(\beta)/SD(\widehat{\beta})$ |
| $\phi^c < \phi^u$ | | | | | |
| 0.04 | 0.15 | NS | 0.638 | 0.651 | 0.957 |
| 0.04 | 0.15 | PS | 0.416 | 0.441 | 0.922 |
| 0.04 | 0.15 | E-PS: linear | 0.335 | 0.361 | 1.002 |
| 0.04 | 0.15 | E-PS: probit | 0.272 | 0.305 | 0.995 |
| 0.04 | 0.15 | E-PS: logit | **0.198** | **0.243** | 0.985 |
| 0.04 | 0.6 | NS | 0.631 | 0.645 | 0.940 |
| 0.04 | 0.6 | PS | 0.425 | 0.451 | 0.889 |
| 0.04 | 0.6 | E-PS: linear | 0.358 | 0.385 | 0.965 |
| 0.04 | 0.6 | E-PS: probit | 0.293 | 0.326 | 0.958 |
| 0.04 | 0.6 | E-PS: logit | **0.217** | **0.260** | 0.959 |
| 0.15 | 0.6 | NS | 0.642 | 0.674 | 0.610 |
| 0.15 | 0.6 | PS | 0.142 | 0.194 | 1.017 |
| 0.15 | 0.6 | E-PS: linear | 0.116 | 0.177 | 1.007 |
| 0.15 | 0.6 | E-PS: probit | 0.085 | 0.158 | 1.018 |
| 0.15 | 0.6 | E-PS: logit | **0.047** | **0.142** | 1.023 |
| $\phi^c > \phi^u$ | | | | | |
| 0.15 | 0.04 | NS | 0.637 | 0.651 | 0.946 |
| 0.15 | 0.04 | PS | 0.122 | 0.174 | 1.039 |
| 0.15 | 0.04 | E-PS: linear | 0.033 | 0.133 | 1.034 |
| 0.15 | 0.04 | E-PS: probit | 0.016 | **0.132** | 1.033 |
| 0.15 | 0.04 | E-PS: logit | **0.002** | 0.134 | 1.031 |
| 0.6 | 0.04 | NS | 0.652 | 0.666 | 0.912 |
| 0.6 | 0.04 | PS | 0.052 | 0.146 | 0.937 |
| 0.6 | 0.04 | E-PS: linear | 0.003 | **0.136** | 0.976 |
| 0.6 | 0.04 | E-PS: probit | **0.000** | 0.137 | 0.978 |
| 0.6 | 0.04 | E-PS: logit | -0.005 | 0.141 | 0.973 |
| 0.6 | 0.15 | NS | 0.644 | 0.667 | 0.713 |
| 0.6 | 0.15 | PS | 0.053 | 0.143 | 0.991 |
| 0.6 | 0.15 | E-PS: linear | 0.012 | 0.129 | 1.053 |
| 0.6 | 0.15 | E-PS: probit | 0.007 | **0.128** | 1.062 |
| 0.6 | 0.15 | E-PS: logit | **0.002** | **0.128** | 1.073 |

Overall, we found that E-PS had the best performance in terms of bias and rMSE across all of the scenarios considered; it was able to substantially reduce spatial confounding bias relative to NS in scenarios where the spatial scale of the confounded variability was smaller than the spatial scale of the unconfounded variability ($\phi^c < \phi^u$) and was able to remove the bias almost entirely when $\phi^c > \phi^u$. In terms of uncertainty quantification, the estimated SE was generally close to the empirical SE, though it slightly overstated the variability in the scenario where ($\phi^c, \phi^u$)=(0.6, 0.15) and slightly understated the variability when ($\phi^c, \phi^u$)=(0.04, 0.6).

In comparing the three variations of E-PS, we found that they performed comparably to each other when $\phi^c > \phi^u$. In contrast, when $\phi^c < \phi^u$ the logistic exposure model (E-PS: logit) was least biased, followed by probit and then linear regression. The superior performance of logistic relative to probit regression was somewhat unexpected, given that the probit exposure model is correctly specified (in terms of the link function). After further examining the results, we observed that the fitted logistic penalized spline exposure model always had greater expected degrees of freedom (EDF) than the corresponding model across all 100 generated datasets in each of the six scenarios considered. Consequently, E-PS logit always adjusted for finer-scale spatial variability than E-PS probit, enabling it to reduce spatial confounding bias to a greater extent.

**Supplemental Appendix C.** Sensitivity analyses for the Moving to Health application

We conducted two sensitivity analyses. First, we explored sensitivity of the findings to the choice of the prior specification for the spatial decay parameter in the NNGP model, where we assumed that $\kappa = 1/\phi$ came from a U(0.1, 30) distribution. This corresponds to an effective range between approximately 0.1 and 36 km, smaller than the effective range in the main result presented in the main text. The following table shows the parameter estimate and 95% posterior credible interval for the smoking and supermarket exposures, under each set of covariate adjustment, for the model using the prior distribution in the main text versus the alternate prior specification. We see that the point estimates are nearly identical, indicating that our results are not sensitive to this choice of prior.

| **Exposure** | **Adjustment** | **Prior** | **Est.** | **2.5%** | **97.5%** |
|---|---|---|---|---|---|
| smoking | base | U(0.1, 30) | 2.431 | 1.366 | 3.310 |
| smoking | base | U(0.02, 5) | 2.433 | 1.372 | 3.314 |
| smoking | ideal | U(0.1, 30) | 3.652 | 2.855 | 4.449 |
| smoking | ideal | U(0.02, 5) | 3.648 | 2.855 | 4.447 |
| smoking | standard | U(0.1, 30) | 2.811 | 1.816 | 3.676 |
| smoking | standard | U(0.02, 5) | 2.811 | 1.805 | 3.677 |
| supermarket | base | U(0.1, 30) | -2.340 | -2.953 | -1.729 |
| supermarket | base | U(0.02, 5) | -2.343 | -2.955 | -1.742 |
| supermarket | ideal | U(0.1, 30) | -3.759 | -4.305 | -3.230 |
| supermarket | ideal | U(0.02, 5) | -3.795 | -4.337 | -3.268 |
| supermarket | standard | U(0.1, 30) | -2.696 | -3.299 | -2.085 |
| supermarket | standard | U(0.02, 5) | -2.710 | -3.303 | -2.101 |

Second, we re-analyzed a subset of the data in which we randomly selected a single individual from each geocoded address having multiple individuals. The following figure shows the results from this sensitivity analysis side-by-side with the original estimates presented in **Figure 5** of the main manuscript. The estimates from this sensitivity analysis tended to be a bit smaller than our original estimates, but the trends across models were very similar to our original analysis and did not change our conclusions.

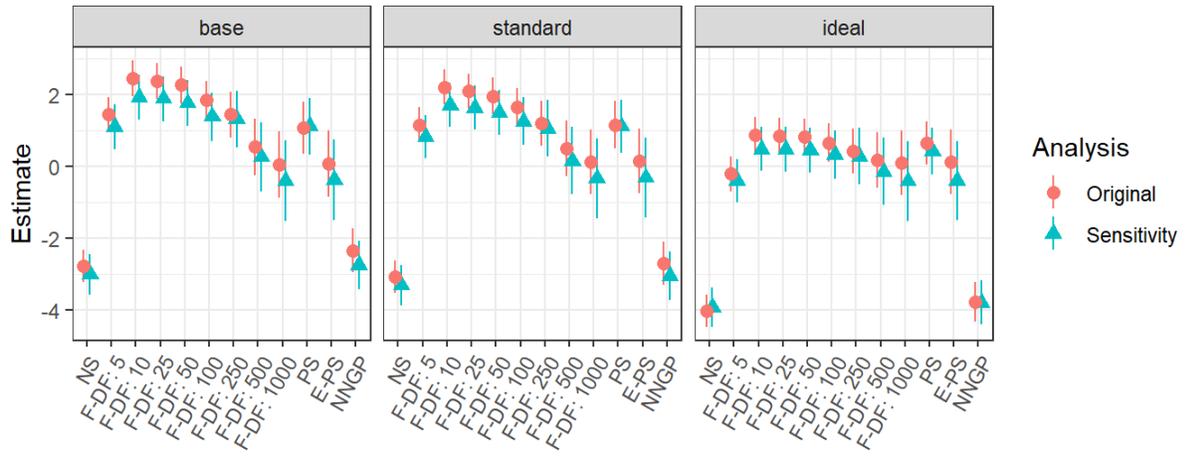
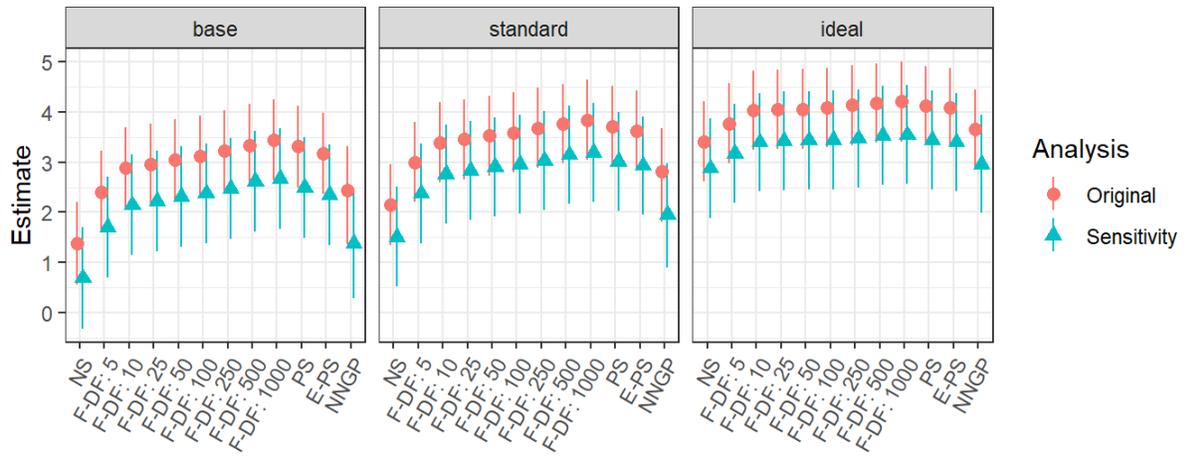

## Supplemental Tables and Figures

**Supplemental Table A.** Computation time (in minutes) across methods for the simulation and application

| Method | No. DF or Basis Dim. ($K$) | Simulation* | | | Application** | | |
| --- | --- | --- | --- | --- | --- | --- | --- |
| | | median | 25% | 75% | Model 1 (base) | Model 2 (standard) | Model 3 (ideal) |
| Fixed DF | 500 | 0.3 | 0.3 | 0.3 | 4.0 | 4.1 | 4.2 |
| Fixed DF | 1000 | 1.0 | 0.9 | 1.0 | 11.0 | 11.2 | 11.3 |
| PS | 500 | 0.4 | 0.4 | 0.5 | | | |
| PS | 1000 | 3.7 | 3.4 | 4.4 | 12.4 | 13.4 | 13.2 |
| E-PS | 500 | 0.7 | 0.7 | 0.8 | | | |
| E-PS | 1000 | 3.9 | 3.6 | 4.2 | 1057/56.0*** | 720/56.4 | 798/58.3 |
| NNGP**** | | 5.7 | 5.7 | 5.8 | 46.5 | 47.4 | 48.9 |

*Results summarized across all scenarios and all simulation repetitions; other methods not shown all took less than 10 seconds

**Other fixed DF methods (not shown) took less than 2 minutes

***Computation time for supermarket/smoking status models (since both exposure variables were included in the outcome regression model, only E-PS has different computation time as a separate exposure model was fit for each exposure variable)

****For simulation, used 10 nearest neighbors and a single processor; for application used 5 nearest neighbors and two parallel processors

**Supplemental Table B.** Change in rMSE versus the non-spatial model. Positive values (**bolded**) indicate that the rMSE is larger under the spatial model

| Method* | $\phi^c$ | $\sigma_x^2 = 0$ | | | $\sigma_x^2 = 0.056$ | | | | $\sigma_x^2 = 0.5$ | | | |
| | | $\phi^u$ | | | | $\phi^u$ | | | | $\phi^u$ | | |
| | | 0.04 | 0.15 | 0.6 | NA | 0.04 | 0.15 | 0.6 | NA | 0.04 | 0.15 | 0.6 |
|---|---|---|---|---|---|---|---|---|---|---|---|---|
| PS | NA | 0.00 | **0.01** | **0.05** | 0.00 | 0.00 | **0.01** | **0.04** | **0.00** | **0.00** | **0.00** | **0.01** |
| PS | 0.04 | **0.00** | 0.55 | 0.84 | 0.00 | -0.04 | **0.34** | **0.57** | -0.08 | -0.11 | -0.04 | 0.00 |
| PS | 0.15 | -0.54 | **0.01** | **0.67** | -0.08 | -0.53 | -0.22 | **0.14** | -0.46 | -0.35 | -0.32 | -0.31 |
| PS | 0.6 | -0.79 | -0.60 | **0.07** | -0.52 | -0.72 | -0.62 | -0.40 | -0.58 | -0.41 | -0.39 | -0.39 |
| E-PS | NA | **0.28** | **1.35** | **8.63** | 0.00 | **0.11** | **0.15** | **0.16** | 0.00 | **0.02** | **0.02** | **0.02** |
| E-PS | 0.04 | **0.03** | **0.93** | **1.00** | -0.58 | -0.40 | -0.18 | -0.17 | -0.45 | -0.31 | -0.25 | -0.24 |
| E-PS | 0.15 | -0.51 | **0.77** | **1.96** | -0.97 | -0.70 | -0.63 | -0.59 | -0.60 | -0.42 | -0.40 | -0.39 |
| E-PS | 0.6 | -0.49 | **0.98** | **6.52** | -1.01 | -0.70 | -0.65 | -0.62 | -0.62 | -0.42 | -0.41 | -0.41 |
| NNGP | NA | 0.00 | 0.00 | **0.01** | 0.00 | **0.00** | 0.00 | **0.01** | **0.00** | **0.00** | **0.00** | **0.00** |
| NNGP | 0.04 | **0.00** | **0.21** | **0.41** | 0.00 | -0.01 | **0.11** | **0.26** | -0.02 | -0.03 | -0.01 | **0.01** |
| NNGP | 0.15 | -0.21 | 0.00 | **0.26** | -0.01 | -0.22 | -0.10 | **0.06** | -0.16 | -0.21 | -0.18 | -0.16 |
| NNGP | 0.6 | -0.37 | -0.25 | **0.02** | -0.04 | -0.39 | -0.29 | -0.15 | -0.30 | -0.28 | -0.25 | -0.24 |

* Results shown are for E-PS and PS with a basis dimension of $K$=1000 (results for these approaches with $K$=500 were very similar)

**Supplemental Table C.** Characteristics of patients from the Moving to Health application across exposure groups being considered

| Variable | Smoking Status | | Supermarket Availability | |
|---|---|---|---|---|
| | current | former | no | yes |
| n | 15,490 | 21,343 | 52,268 | 65,597 |
| age (mean (SD)) | 40.8 (12.9) | 46.0 (12.3) | 42.2 (13.7) | 40.3 (13.1) |
| height (mean (SD)) | 67.2 (3.9) | 67.1 (3.8) | 66.5 (4.0) | 66.5 (4.0) |
| Male (%) | 49.8 | 46.3 | 39.9 | 39.2 |
| Medicaid (%) | 3.7 | 3.6 | 3.6 | 3.6 |
| Race/ethnicity (%) | | | | |
|   Other/Unknown | 6.1 | 4.4 | 4.2 | 4.4 |
|   Asian | 67.6 | 75.1 | 66.3 | 66.8 |
|   Black | 9.7 | 5.3 | 6.7 | 7.5 |
|   Hispanic | 6.2 | 5.8 | 6.2 | 6.4 |
|   White | 10.3 | 9.4 | 16.5 | 14.9 |
| Property value (in $1,000s; median [IQR]) | 254.2 [152.9, 358.7] | 303.4 [198.8, 422.4] | 323.4 [232.4, 463.7] | 290.0 [168.9, 412.0] |
| Weight (pounds; mean (SD)) | 182.6 (47.5) | 185.8 (47.8) | 176.5 (47.5) | 173.6 (46.4) |
| BMI category (%) | | | | |
|   <18.5 | 1.5 | 0.7 | 1.4 | 1.5 |
|   [18.5, 25) | 33.0 | 29.5 | 36.6 | 40.2 |
|   [25, 30) | 33.1 | 35.1 | 32.5 | 31.8 |
|   [30, 35) | 18.0 | 18.8 | 16.7 | 14.8 |
|   ≥35 | 14.4 | 15.9 | 12.8 | 11.7 |

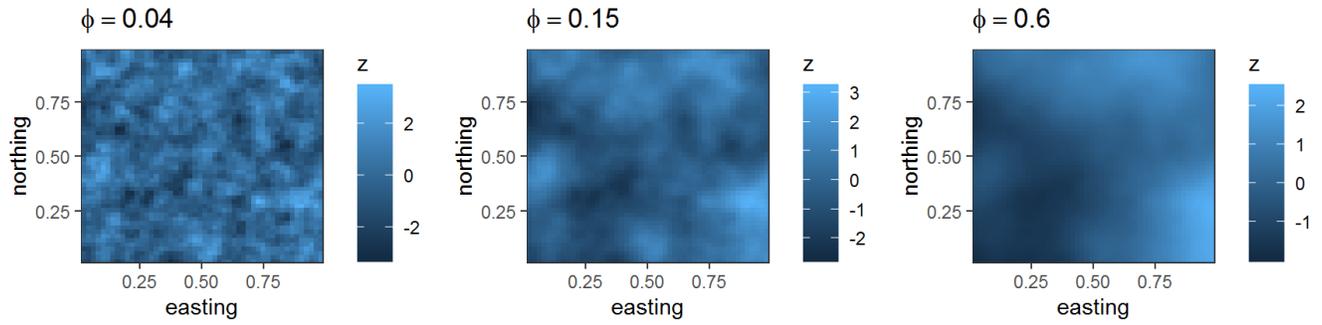

**Supplemental Figure A.** Realization of Gaussian process used to generate the spatial components $z(s)$ for the different values of $\phi$ considered.

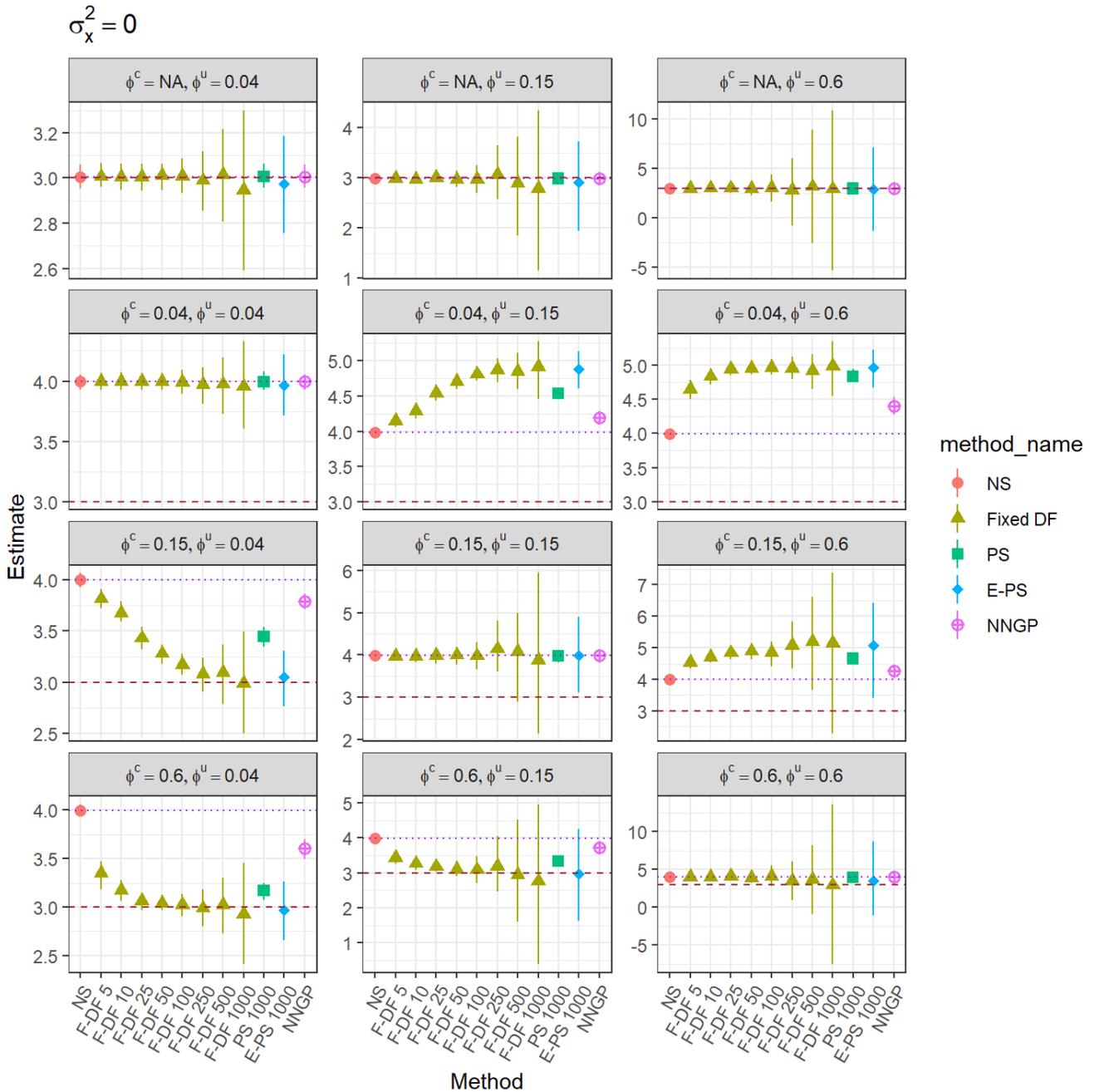

**Supplemental Figure B.** Mean (interquartile range) of $\beta$ parameter estimates for the setting of no non-spatial variability in the exposure ($\sigma_x^2 = 0$) across different spatial scales of the unconfounded and confounded variability ($\phi^u$ and $\phi^c$, respectively). Dashed horizontal line corresponds to the true value of $\beta$; dotted horizontal line corresponds to the mean estimate from the non-spatial model. Note that for this setting with $\sigma_x^2 = 0$ there is no scenario without an independent spatial component ($\phi^u = NA$), because then there would be no unconfounded variability in $x$ (i.e., $x$ would completely colinear with the confounder $\tilde{z}_i^c$). NS, non-spatial; F-DF, fixed degrees of freedom; PS, penalized spline; E-PS, exposure-penalized spline; NNGP, nearest neighbor Gaussian process.

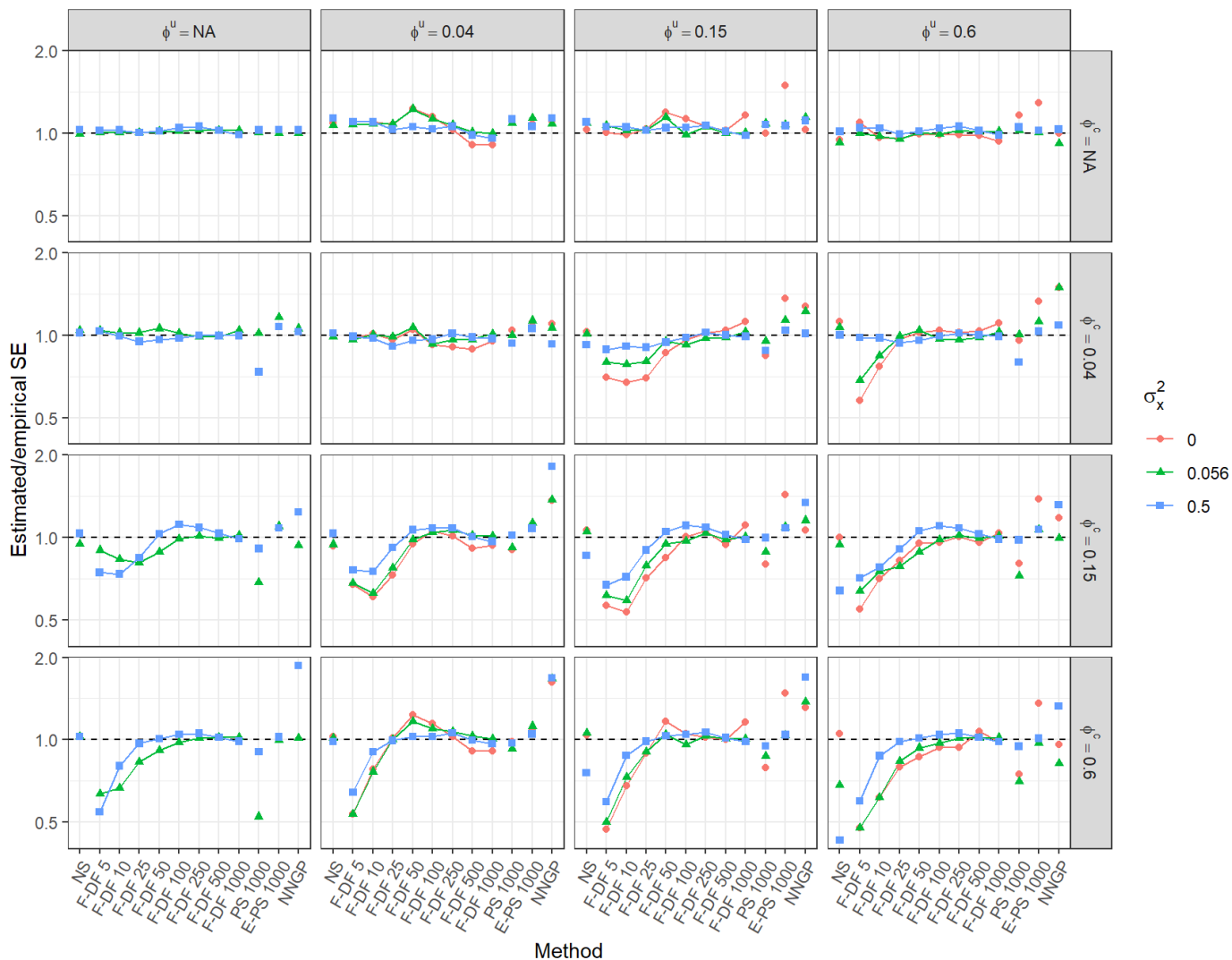

**Supplemental Figure C.** Ratio of the estimated standard error (or posterior standard deviation) and the empirical standard error (calculated as the SD of the parameter estimates across Monte Carlo iterations). Results shown across different values of the proportion of non-spatial variability in exposure (parameterized by $\sigma_x^2$) and different spatial scales of the unconfounded and confounded variability ($\phi^u$ and $\phi^c$, respectively). NS, non-spatial; F-DF, fixed degrees of freedom; PS, penalized spline; E-PS, exposure-penalized spline; NNGP, nearest neighbor Gaussian process.

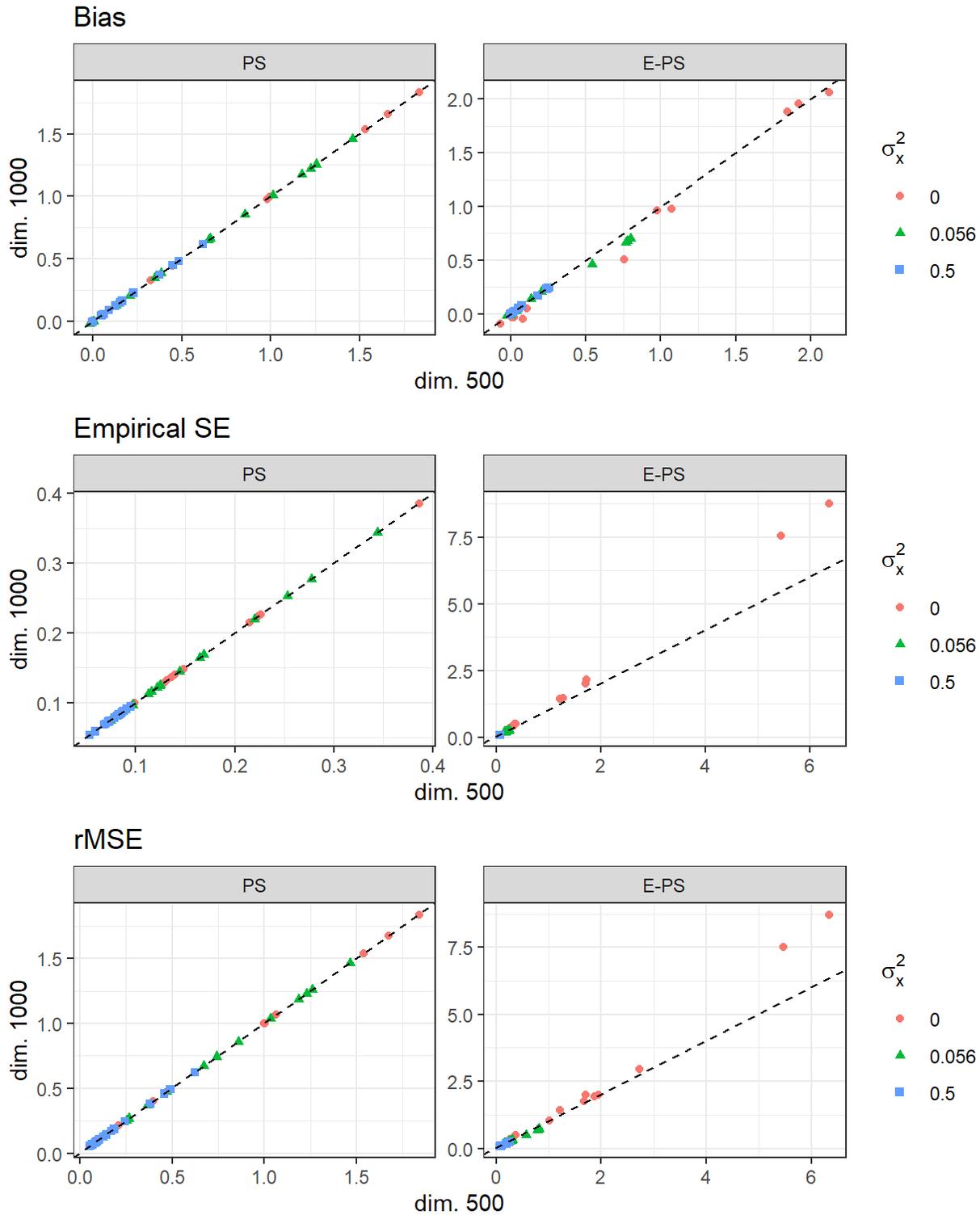

**Supplemental Figure D.** Comparison of bias, empirical standard error (SE), and root mean squared error (rMSE) under the penalized spline (PS) and exposure-penalized spline (E-PS) approaches when the basis dimension ($K$) is 500 versus 1000.